\documentclass[aip,
amsmath,amssymb,
preprint,%
$reprint,%
numerical,
]{revtex4-1}
\usepackage{graphicx}
\usepackage{dcolumn}
\usepackage{bm}
\usepackage{hyperref}
\usepackage[utf8]{inputenc}
\usepackage[T1]{fontenc}
\usepackage{mathptmx}
\usepackage{etoolbox}
\makeatletter
\def\@email#1#2{%
	\endgroup
	\patchcmd{\titleblock@produce}
	{\frontmatter@RRAPformat}
	{\frontmatter@RRAPformat{\produce@RRAP{*#1\href{mailto:#2}{#2}}}\frontmatter@RRAPformat}
	{}{}
}%
\makeatother
\begin{document}

	\title{Effect of rigid top surface on the onset of phototactic bioconvection}
	\author{S. K. Rajput}
	\affiliation{ 
		Department of Mathematics, PDPM Indian Institute of Information Technology Design and Manufacturing,
		Jabalpur 482005, India.
	}%
	
	

	\begin{abstract}
		Phototaxis is the movement of microorganisms towards or away from light. In presence of dim (intense) light, microorganisms move towards (away from) the light sources, which is known as positive (negative) phototaxis. In this study, we present a model in which isotropic scattering algae suspension is illuminated by both diffuse and collimated irradiation, and the top surface is assumed to be rigid. Bimodal steady states result from the scattering Effect (for almost purely scattering suspension), which transits into a unimodal steady state as diffuse light intensity increases. The linear stability of the same suspension is also examined using the linear perturbation theory, which predicts the solutions' stable and oscillatory nature for specific parameter values. Here, we also observe that suspension becomes more stable in a rigid top surface and diffuse irradiation.
		
	\end{abstract}
	
	\maketitle

	\section{INTRODUCTION}
	The phenomenon known as $bioconvection$ is the macroscopic convective motion of a fluid containing self-propelled motile microorganisms. These microorganisms are denser than the medium (water here) in which they swim upward on average. Mostly algae and bacteria show this type of behavior. When the microorganisms stop swimming, the pattern formation in bioconvection disappears. However, upswimming and higher density are not necessary for pattern formation. There are examples where both upswimming and higher density are not involved in pattern formation. Swimming microorganisms change their orientation in response to different types of environmental stimuli. These responses are known as $taxes$. $Gravitaxis$, $chemotaxis$, $phototaxis$, and $gyrotaxis$ are the most important examples of $taxes$. $Gravitaxis$ is the swimming response to gravitational acceleration; $chemotaxis$ is the swimming response due to chemicals; $gyrotaxis$ is caused by the balance between a couple of torques due to gravity and local shear flow; and  $phototaxis$ is defined as the swimming response toward the light source (positive phototaxis) or away from the light source (negative phototaxis). Self-shading is the mechanism by which microorganisms absorb the light incident on them from directly above and produce shadow below them. This article considers the Effect of phototaxis only.
	
	Experimental studies have shown that pattern formation in bioconvection may be significantly influenced by various forms of illumination intensity (such as diffuse irradiation)~\cite{1wager1911,2kitsunezaki2007}. 
	Strong (bright) light damages the stable patterns or prevents the formation of patterns in a suspension of motile microorganisms in well-stirred culture.
	The pattern's size, shape, structure, and symmetry may all be affected by the light intensity~\cite{3kessler1985,4williams2011,5kessler1989}. Variations in bioconvection patterns caused by light intensity can be explained through the following circumstances.  First, the phototactic algae obtain energy via the process of photosynthesis. Consequently, the behavioral response modifies their swimming behavior (phototaxis).
	When the light intensity $G$ is less than or larger than its critical value $G_c$, cells swim toward the light source (positive phototaxis) and away from it (negative phototaxis). Thus, the algae cells try to accumulate at a suitable location where they can find optimal light intensity ($G=G_c$). The second possible reason for pattern modifications may be the absorption and scattering of light. Due to the absorption of light by the cells, the light intensity decreases along the incident path and due to scattering; first, light is deflected away from the incident path, which causes the intensity to decrease, and then light scatter from another point and reaches that point, which causes the intensity to increases~\cite{7ghorai2010}. The diffuse (scattered) irradiation illuminates the suspension more uniformly. So, under diffuse irradiation algae cells move incompletely in their environment. As a result, self-shading is less effective under diffuse irradiation. Therefore, diffuse irradiation may also affect pattern formation in bioconvection.\par
	
	In the present study, we use the phototaxis model provided by Panda $et$ $al$.~\cite{15panda2016}, which utilized the Navier-Stokes equations for an incompressible fluid combined with a conservation equation for microorganisms and radiative transfer equation (RTE) for light transport. Due to the presence of clouds (water droplets) and dust particles, the radiation from the sun gets diffuse (scattered), and diffuse irradiation may affect the accumulation rate of algae cells during the convection. In recent years, green algae that are primarily phototactic are employed for carbon dioxide fixation through photosynthesis in photo-bioreactors, and biofuels may be made from the produced biomass. Bioconvection can be less worthy in algal biofuel production. Therefore, while building effective photo-bioreactors, it is essential to consider the Effect of diffuse irradiation on phototaxis (photosynthesis) and the resulting bioconvection, especially for dense algal suspensions. Due to their requirement to photosynthesize, many motile algae are highly phototactic in their natural ecosystems. Therefore, to fully characterize their swimming behavior, a realistic phototaxis model should consider the effects of collimated irradiation with diffuse irradiation.\par
	
	Let the suspension be illuminated from the top by diffuse irradiation. The equilibrium state for such suspension is formed when the fluid velocity becomes zero. The vertical swimming, driven by positive and negative phototaxis, is balanced by the diffusion given by random cell movement. As a result, the concentrated sublayer is formed and its position is determined by the critical intensity $G_c$. A sublayer is formed on top (bottom) of the suspension if the intensity throughout the suspension is higher (lower) than the critical intensity, but if the critical intensity $G_c$ is between the maximum and minimum value of the light intensity, a sublayer forms in between the suspension's top and bottom boundaries. Above (below), the sublayer is a gravitationally stable (unstable) region. If the fluid layer becomes unstable, the fluid flows from the lower unstable region and penetrates the top stable region. This is an example of penetrative bioconvection~\cite{9straughan1993,10ghorai2005,11panda2016}.
	
	\begin{figure}[!h]
		\centering
		\includegraphics[width=14cm]{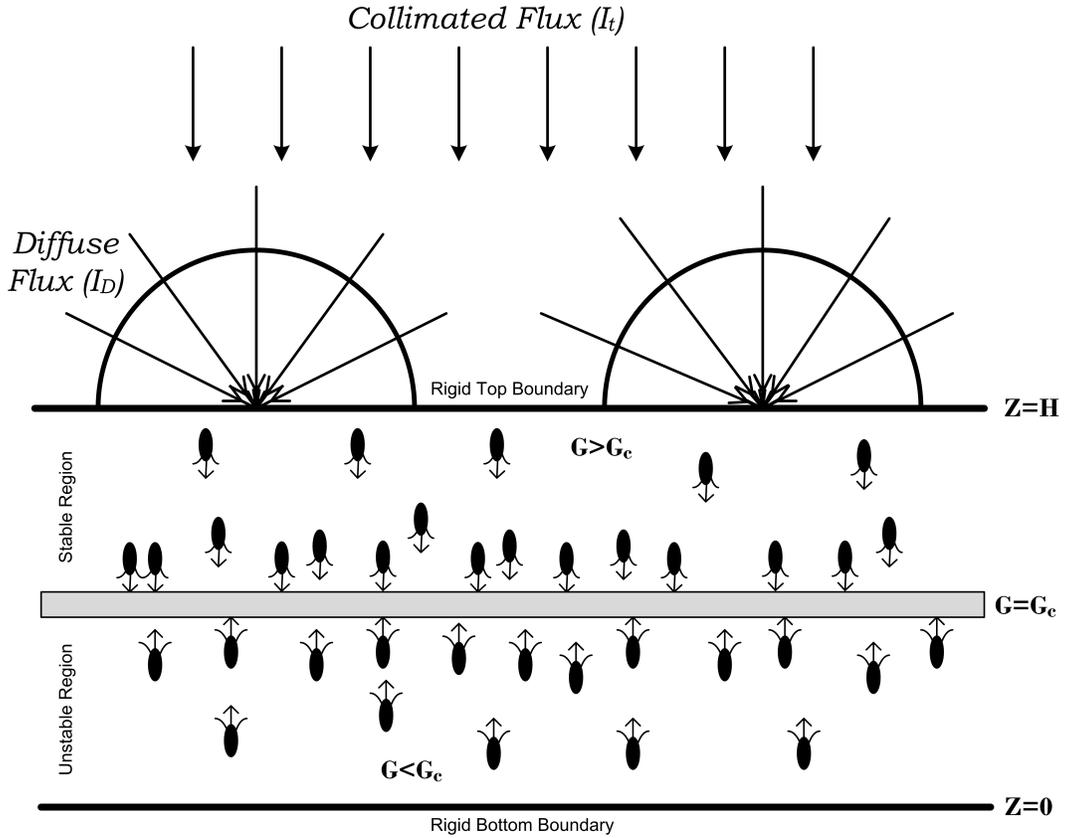}
		\caption{\footnotesize{Formation of the sublayer at $G=G_c$ in the interior of the algal suspension, where $G_c$ is the critical light intensity. Above (below) the sublayer, the suspension is stable (unstable).}}
		\label{fig1}
	\end{figure}
	
	There have been done lots of work on phototactic bioconvection. Vincent and Hill~\cite{12vincent1996} attempted to simulate proprietary phototactic bioconvection (i.e., no orientation bias due to gravity) for algae suspension in a shallow layer. They constructed a simple top-down photoresponse based on light intensity. They believe that the main Effect of the cells on the suspension is due to their negative buoyancy, and other sources of overall stress have been overlooked. Vincent and Hill~\cite{12vincent1996} used the well-known Beer-Lambert law to simulate light intensity. They performed linear stability analysis and found the disturbance's stationary and oscillatory nature. Ghorai and Hill~\cite{10ghorai2005} quantified phototactic bioconvection in two dimensions where the suspension is confined between a rigid bottom and stress-free upper and side walls. They also address the significant inaccuracies in Vincent and Hill's equilibrium solution. Ghorai $et$ $al$.
	~\cite{7ghorai2010} studied the effects of light scattering and found bimodal equilibrium-state configurations for almost purely scattering suspension. Ghorai and Panda~\cite{13ghorai2013} used linear theory to study the initiation of bioconvection in an anisotropic scattering solution of phototactic algae emphasizing the forward scattering Effect, and they found a considerable impact on solutions by forward light scattering. Panda and Ghorai~\cite{14panda2013} used linear simulations to simulate 2-D phototactic bioconvection in an isotropic scattering suspension. Due to the scattering Effect, the bioconvective patterns found in their investigation were qualitatively different from those reported by Ghorai and Hill~\cite{10ghorai2005} at higher critical wavelengths. Panda and Singh~\cite{11panda2016} numerically investigated the linear stability of phototactic bioconvection in 2-D using Vincent and Hill's continuum model. They noticed a considerable stabilizing effect on the suspension due to rigid side walls. In previous studies, diffuse irradiation's effects were not considered.
	In this direction, Panda $et$ $al$.~\cite{15panda2016} proposed a model in which they looked at how the diffuse radiation affected an isotropic scattering algal suspension. They found that diffuse irradiation strongly stabilizes isotropically scattering suspensions, and diffuse irradiation has a stronger effect on the critical state ($k_c$, $R_c$) compared with collimated irradiation alone. Panda~\cite{8panda2020} studied the impact of forward anisotropic scattering on the onset of phototactic bioconvection using both diffuse and collimated irradiation. Panda $et$ $al$.~\cite{16panda2022} studied the Effect of oblique irradiation on algal suspensions and observed that the location of the maximum concentration of microorganisms in a suspension shifts towards the top of the suspension and the value of the maximum concentration increases as the angle of incidence increases. In the same article, they used Snell's law for the refraction of light and checked the linear stability of the suspension for different angles of incidence. However, they neglected the scattering Effect in their study. Recently, Kumar~\cite{17kumar2022} investigated the Effect of oblique collimated irradiation on the isotropic scattering algal suspension. He found both types of solutions (stationary and overstable) for certain ranges of parameters. More recently, Kumar studied the Effect of collimated irradiation on the algae suspension where both vertical walls were assumed to be rigid. In their study, he found the stabilizing Effect on the suspension due to rigid walls. Inspired by Kumar's recent study, we investigate the Effect of rigid vertical walls on the algal suspension where suspension is illuminated by both collimated and diffuse irradiation.\par
	
	This article is structured as follows for further investigation:  first, we propose a mathematical formulation of the phototaxis model and compute the basic equilibrium state. After that, we establish the linear stability of the basic equilibrium state based on the linear perturbation theory and solve it numerically by Newton-Raphson-Kantorovich (NRK) method and find the neutral curves. Finally, the results are discussed by using these neutral curves. 
	
	\section{MATHEMATICAL FORMULATION}
	
	Consider the movement in a dilute phototactic algal suspension within a layer of finite depth $H$ and infinite lateral extent. In this model, the suspension is illuminated from the top by collimated and diffuse irradiation (see Fig.\ref{fig1}). Let $I(\boldsymbol{x},\boldsymbol{s})$ denote the intensity of light present at a given location $\boldsymbol{x}$ in the unit direction $\boldsymbol{s}=\cos{\theta}\hat{z}+\sin{\theta}(\cos{\phi}\hat{x}+\sin{\phi}\hat{y})$, where $\boldsymbol{x}$ is measured relative to a rectangular cartesian coordinate system o(x,y,z) with the z-axis vertically up and $\hat{x}$, $\hat{y}$, $\hat{z}$ are unit vectors along the x, y, z axes. 

	\subsection{\label{sec:level3}PHOTOTAXIS WITH ABSORPTION AND SCATTERING}
	
	The light intensity for an absorbing and scattering medium is calculated by RTE, which is stated as 
	\begin{equation}\label{1}
		\frac{dI(\boldsymbol{x},\boldsymbol{s})}{ds}+(a+\sigma_s)I(\boldsymbol{x},\boldsymbol{s})=\frac{\sigma_s}{4\pi}\int_{0}^{4\pi}I(\boldsymbol{x},\boldsymbol{s'})f(\boldsymbol{s},\boldsymbol{s'})d\Omega',
	\end{equation}
	where $\alpha,\sigma_s$ stand for the absorption and scattering coefficients respectively, and $f(\boldsymbol{s},\boldsymbol{s'})$ is the scattering phase function, which provides the angular distribution of the light intensity scattered from $\boldsymbol{s'}$ direction into $\boldsymbol{s}$ direction. Here, the algae cells assume the light is scattered isotropically. So for simplicity, we use $f(\boldsymbol{s},\boldsymbol{s'})=1$~\cite{15panda2016} in further calculations.

	\begin{figure}[!h]
		\centering
		\includegraphics[width=14cm]{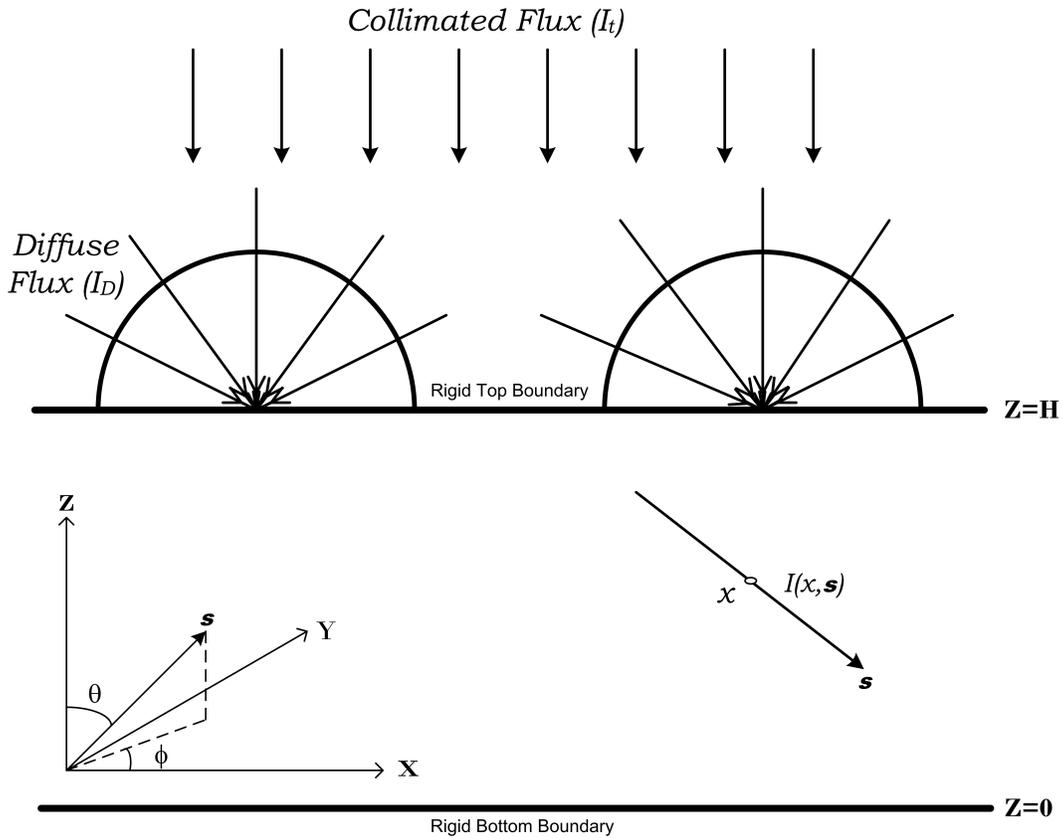}
		\caption{\footnotesize{Geometric configuration of the problem.}}
		\label{fig2}
	\end{figure}

	We assume  that the suspension's top boundary is diffusive, then the intensity on the top of the suspension is thus given by
	\begin{equation*}
		I(\boldsymbol{x}_b,\boldsymbol{s})=I_t\delta(\boldsymbol{s}-\boldsymbol{s_0})+\frac{I_D}{\pi}, 
	\end{equation*}
	where $\boldsymbol{x}_b=(x,y,H)$ is the location on the top boundary surface. Here, $I_t$ and $I_D$ are the magnitudes of collimated and diffuse irradiation respectively~\cite{8panda2020,15panda2016}. If we take $I_D=0$, this model will be similar to the previous model proposed by Ghorai $et$ $al$. Now, we assume that absorption and scattering coefficients are proportional to the concentration of cells $n$. So $a=\alpha n(\boldsymbol{x})$ and $\sigma_s=\beta n(\boldsymbol{ x})$, then the RTE becomes
	\begin{equation}\label{2}
		\frac{dI(\boldsymbol{x},\boldsymbol{s})}{ds}+(\alpha+\beta)nI(\boldsymbol{x},\boldsymbol{s})=\frac{\beta n}{4\pi}\int_{0}^{4\pi}I(\boldsymbol{x},\boldsymbol{s'})d\Omega'.
	\end{equation}
	In the medium, the total intensity at a fixed point $\boldsymbol{x}$ is 
	\begin{equation*}
		G(\boldsymbol{x})=\int_0^{4\pi}I(\boldsymbol{x},\boldsymbol{s})d\Omega,
	\end{equation*}
	and the radiative heat flux is given by
	\begin{equation}\label{3}
		\boldsymbol{q}(\boldsymbol{x})=\int_0^{4\pi}I(\boldsymbol{x},\boldsymbol{s})\boldsymbol{s}d\Omega.
	\end{equation}
	Let $\boldsymbol{p}$ be the unit vector in the cell swimming direction and $<\boldsymbol{p}>$ is the mean swimming direction.
	For many species of microorganisms, the swimming speed is free from the light condition, position, time, and direction~\cite{18hill1997}. Here, we assumed that the cells and the fluid flow at the same speed, then the average cell
	swimming velocity is defined as
	\begin{equation*}
		\boldsymbol{W}_c=W_c<\boldsymbol{p}>,
	\end{equation*}
	where $W_c$ is the average cell swimming speed and the cell's mean swimming direction $<\boldsymbol{p}>$ is calculated by
	\begin{equation}\label{4}
		<\boldsymbol{p}>=-T(G)\frac{\boldsymbol{q}}{\varpi+|\boldsymbol{q}|},
	\end{equation}
	where $\varpi$ is a non-negative constant and $T(G), $ is the taxis response function (taxis function), which shows the response of algae cells to light and has the mathematical form such that 
	\begin{equation*}
		T(G)=\left\{\begin{array}{ll}\geq 0, & \mbox{if } G(\boldsymbol{x})\leq G_{c}, \\
			< 0, & \mbox{if }G(\boldsymbol{x})>G_{c}.  \end{array}\right. 
	\end{equation*}
	
	The mean swimming direction becomes zero at the critical light intensity ( $G = G_c)$.
	Generally, the exact functional form of taxis function depends on the species of the microorganisms~\cite{12vincent1996}. Since the source of light intensity lies in the opposite direction of the intensity flux vector, the negative sign arises in Eq.~(\ref{4}). The non-negative $\varpi$ is introduced to address the problem of isotropic lighting conditions, but the light intensity across the suspension is not isotropic, so we use $\varpi=0$~\cite{7ghorai2010} here. Thus, the mean swimming direction is calculated using $\varpi= 0$ in Eq.~(\ref4).

	\subsection{GOVERNING EQUATIONS}
	
	We assume a monodisperse cell population that can be modeled by a continuous
	distribution reminiscent of the previous models on bioconvection. The algal suspension is dilute $(nv<<1)$ so that the volume fraction of the cells is small and cell-cell interactions are negligible. Each cell has volume $v$ and density $\rho+\Delta\rho$, with $\rho$ being the water density $(\Delta\rho<<\rho)$. In elemental volume, the average fluid velocity is $\boldsymbol{u}$, and the number of algal cells is $n$. Suppose that the
	suspension is incompressible. Then the conservation equation for the total mass is
	\begin{equation}\label{5}
		\boldsymbol{\nabla}\cdot \boldsymbol{u}=0.
	\end{equation}
	For simplicity, we assume that Stokeslets dominate the Effect of cells on the suspension except for their negative buoyancy,  and all other effects on the bulk stress are minor and can be ignored.
	So, the momentum equation under the Boussinesq approximation 
	\begin{equation}\label{6}
		\rho\frac{D\boldsymbol{u}}{Dt}=-\boldsymbol{\nabla} P_e+\mu\nabla^2\boldsymbol{u}-nvg\Delta\rho\hat{\boldsymbol{z}},
	\end{equation}
	where $D/Dt=\partial/\partial t+\boldsymbol{u}\cdot\boldsymbol{\nabla}$ is the material or total derivative, $P_e$ is the excess pressure above hydrostatic, $\mu$ is the viscosity of the suspension which is assumed to be that of fluid and $\hat{z}$ is the unit vector in a vertically upward direction.\newline 
	The equation of cell conservation is given by
	\begin{equation}\label{7}
		\frac{\partial n}{\partial t}=-\boldsymbol{\nabla}\cdot \boldsymbol{F},
	\end{equation}
	where $\boldsymbol{F}$ is the total cell flux which is given by
	\begin{equation}\label{8}
		\boldsymbol{F}=nu+nW_c<\boldsymbol{p}>-\boldsymbol{D}\boldsymbol{\nabla} n.
	\end{equation}
	The first term on R.H.S. of Eq.~(\ref{8}) is flux due to the advection of cells by the bulk fluid flow, the second term arises due to the average swimming of the cells and the third term represents the random component of the cell locomotion. The diffusivity tensor $\boldsymbol{D}$ is assumed to be isotropic and constant, with the result that $\boldsymbol{D} = DI$. The expression for the cell flux vector in Eq.~(\ref{8}) has two key assumptions. First, cells are assumed to be purely phototactic, so the Effect of viscous torque, which may contribute to the cell swimming direction, is ignored. Second, a constant value for the diffusion tensor is assumed, which should be determined from the swimming velocity autocorrelation function. Making these assumptions may remove the Fokker-Planck equation from the governing equations. The resultant model can be used as a suitable limiting case to estimate the problem's difficulty before constructing a more complex model.
	
	\subsection{BOUNDARY CONDITIONS}
	When an upper boundary is open to the air, cells often collect at the top of the suspension and form a rigid packed layer. This is a practical example of the upper layer being rigid. Therefore, in this model, lower and upper boundaries are considered rigid. Here, the normal and tangential component of fluid velocity becomes zero at the boundaries, and there is no flow of cells through them.
	Thus, the boundary conditions are
	\begin{equation}\label{9}
		\boldsymbol{u}\cdot\hat{\boldsymbol{z}}=0\qquad on\quad z=0,H,
	\end{equation}
	\begin{equation}\label{10}
		\boldsymbol{F}\cdot\hat{\boldsymbol{z}}=0\qquad on\quad z=0,H.
	\end{equation}
	For rigid boundaries,
	\begin{equation}\label{11}
		\boldsymbol{u}\times\hat{\boldsymbol{z}}=0\qquad on\quad z=0,H.
	\end{equation}
	
	We assume that the top boundary is exposed to the uniform diffuse irradiation, then the boundary condition for intensities are
	\begin{subequations}
		\begin{equation}\label{12a}
			I(x, y, z = 0, \theta, \phi)=I_t\delta(\boldsymbol{s}-\boldsymbol{s_0})+\frac{I_D}{\pi},\quad (\pi/2\leq\theta\leq\pi),
		\end{equation}
		\begin{equation}\label{12b}
			I(x, y, z = 0, \theta, \phi) =0,\quad (0\leq\theta\leq\pi/2).
		\end{equation}
	\end{subequations}

	\subsection{DIMENSIONLESS EQUATIONS}
	The governing equations are made dimensionless
	by choosing $H$ as length-scale, $H^2/D$ as time scale, $D/H$ as velocity scale, $\mu D/H^2$ as pressure scale, and $\Bar{n}$ as concentration scale, which are given below 
	\begin{equation}\label{13}
		\boldsymbol{\nabla}\cdot\boldsymbol{u}=0,
	\end{equation}
	\begin{equation}\label{14}
		S_{c}^{-1}\left(\frac{D\boldsymbol{u}}{Dt}\right)=-\nabla P_{e}-Rn\hat{\boldsymbol{z}}+\nabla^{2}\boldsymbol{u},
	\end{equation}
	\begin{equation}\label{15}
		\frac{\partial{n}}{\partial{t}}=-{\boldsymbol{\nabla}}\cdot{\boldsymbol{F}},
	\end{equation}
	where
	\begin{equation}\label{16}
		{\boldsymbol{F}}=n{\boldsymbol{u}}+nV_{c}<{\boldsymbol{p}}>-{\boldsymbol{\nabla}}n.
	\end{equation}

	Here $S_{c}=\nu/{D}$ is the Schmidt number, $V_c=W_cH/D$ is scaled swimming speed, and $R=\overline{n}\vartheta g\Delta{\rho}H^{3}/\nu\rho{D}$ is the Rayleigh number (a control parameter).
	In dimensionless form, the boundary conditions become
	\begin{equation}\label{17}
		\boldsymbol{u}\cdot\hat{\boldsymbol{z}}=0\qquad on\quad z=0,1,
	\end{equation}
	\begin{equation}\label{18}
		\boldsymbol{F}\cdot\hat{\boldsymbol{z}}=0\qquad on\quad z=0,1.
	\end{equation}
	For rigid boundaries,
	\begin{equation}\label{19}
		\boldsymbol{u}\times\hat{\boldsymbol{z}}=0\qquad on\quad z=0,1.
	\end{equation}

	The RTE in dimensionless form is
	\begin{equation}\label{20}
		\frac{dI(\boldsymbol{x},\boldsymbol{s})}{ds}+\kappa nI(\boldsymbol{x},\boldsymbol{s})=\frac{\sigma n}{4\pi}\int_{0}^{4\pi}I(\boldsymbol{x},\boldsymbol{s'})d\Omega',
	\end{equation}
	where $\kappa=(\alpha+\beta)\Bar{n}H$, $\sigma=\beta\Bar{n}H$ are the non-dimensional extinction coefficient and scattering coefficient respectively. The scattering albedo $\omega=\sigma/\kappa$ measures the scattering efficiency of microorganisms. In terms of scattering albedo $\omega$, Eq.~(\ref{20}) can be written as
	\begin{equation}\label{21}
		\frac{dI(\boldsymbol{x},\boldsymbol{s})}{ds}+\kappa nI(\boldsymbol{x},\boldsymbol{s})=\frac{\omega\kappa n}{4\pi}\int_{0}^{4\pi}I(\boldsymbol{x},\boldsymbol{s'})d\Omega',
	\end{equation}
	where the value of the scattering albedo $\omega\in$ [0, 1] and $\omega=1 (\omega=0)$ implies purely scattering (purely absorbing) medium. In the form of direction cosine, RTE becomes,
	
	\begin{equation}\label{22}
		\xi\frac{dI}{dx}+\eta\frac{dI}{dy}+\nu\frac{dI}{dz}+\kappa nI(\boldsymbol{x},\boldsymbol{s})=\frac{\omega\kappa n}{4\pi}\int_{0}^{4\pi}I(\boldsymbol{x},\boldsymbol{s'})d\Omega',
	\end{equation}
	where $\xi,\eta$ and $\nu$ are the direction cosines in x, y and z direction. In dimensionless form, the intensity at boundaries becomes,
	\begin{subequations}
		\begin{equation}\label{23a}
			I(x, y, z = 1, \theta, \phi)=I_t\delta(\boldsymbol{s}-\boldsymbol{s_0})+\frac{I_D}{\pi} ,\qquad (\pi/2\leq\theta\leq\pi),
		\end{equation}
		\begin{equation}\label{23b}
			I(x, y, z = 0, \theta, \phi) =0,\qquad (0\leq\theta\leq\pi/2). 
		\end{equation}
	\end{subequations}

	\section{THE BASIC (EQUILIBRIUM) STATE SOLUTION}
	
	Equations $(\ref{13})-(\ref{16})$ and $(\ref{2})$ have an equilibrium solution in the form of
	
	\begin{equation}\label{24}
		\boldsymbol{u}=0,~~~n=n_s(z)\quad and\quad  I=I_s(z,\theta).
	\end{equation}
	Therefore, the total intensity $G_s$ and radiative flux $\boldsymbol{q}_s$ at the equilibrium state are given by
	
	\begin{equation*}
		G_s=\int_0^{4\pi}I_s(z,\theta)d\Omega,\quad 
		\boldsymbol{q}_s=\int_0^{4\pi}I_s(z,\theta)\boldsymbol{s}d\Omega,
	\end{equation*}
	and the governing equation for $I_s$ can be written as
	\begin{equation}\label{25}
		\frac{dI_s}{dz}+\frac{\kappa n_sI_s}{\nu}=\frac{\omega\kappa n_s}{4\pi\nu}G_s(z).
	\end{equation}
	
	We decompose the basic state intensity into collimated part $I_s^c$ and diffuse part $I_s^d$ ( which occurs due to scattering) such that $I_s=I_s^c+I_s^d$. The equation governs the collimated part $I_s^c$. 
	
	\begin{equation}\label{26}
		\frac{dI_s^c}{dz}+\frac{\kappa n_sI_s^c}{\nu}=0,
	\end{equation}
	
	subject to the boundary conditions

	\begin{equation}\label{27}
		I_s^c( 1, \theta) =I_t\delta(\boldsymbol{s}-\boldsymbol{ s}_0),\qquad (\pi/2\leq\theta\leq\pi), 
	\end{equation}
	Now $I_s^c$ is given by
	\begin{equation}\label{28}
		I_s^c=I_t\exp\left(\int_z^1\frac{\kappa n_s(z')}{\nu}dz'\right)\delta(\boldsymbol{s}-\boldsymbol{s_0}), 
	\end{equation}
	
	and the diffused part is governed by  
	\begin{equation}\label{29}
		\frac{dI_s^d}{dz}+\frac{\kappa n_sI_s^d}{\nu}=\frac{\omega\kappa n_s}{4\pi\nu}G_s(z),
	\end{equation}
	subject to the boundary conditions
	\begin{subequations}
		\begin{equation}\label{30a}
			I_s^d( 1, \theta) =\frac{I_D}{\pi},\qquad (\pi/2\leq\theta\leq\pi), 
		\end{equation}
		\begin{equation}\label{30b}
			I_s^d( 0, \theta) =0,\qquad (0\leq\theta\leq\pi/2). 
		\end{equation}
	\end{subequations}
	
	Eq.~(\ref{30a}) can be justified on the assumption that the incident radiation is diffuse on the azimuthal symmetric top surface ($z=1$) of the suspension and thus, $I_s^d(1,\theta)$, $\pi/2\leq\theta\leq \pi$, is assumed to be direction independent and hence constant. Hence, the magnitude of the diffuse irradiation, $I_D$ reduces to
	
	\begin{equation*}
		I_D=\pi I_s^d(1,\theta)
	\end{equation*}

	Now the total intensity, $G_s=G_s^c+G_s^d$ in the equilibrium state can be written as
	\begin{equation}\label{31}
		G_s^c=\int_0^{4\pi}I_s^c(z,\theta)d\Omega=I_t\exp\left(-\int_z^1\kappa n_s(z')dz'\right),
	\end{equation}
	\begin{equation}\label{32}
		G_s^d=\int_0^{\pi}I_s^d(z,\theta)d\Omega.
	\end{equation}
	
	We get the well-known Lambert-Beer law $G_s=G_s^c$ for no scattering. If we define the optical thickness as 
	\begin{equation*}
		\tau=\int_z^1 \kappa n_s(z')dz',
	\end{equation*}
	
	then the total intensity $G_s$ becomes a function of $\tau$ only. Further, the non-dimensional total intensity, $\Lambda(\tau)=G_s(\tau)/I_t$, satisfies the following Fredholm Integral Equation (FIE),
	
	\begin{equation}\label{33}
		\Lambda(\tau) = \frac{\omega}{2}\int_0^\kappa \Lambda(\tau')E_1(|\tau-\tau'|)d\tau'+e^{-\tau}+2I_DE_2(\tau),
	\end{equation}
	
	Where $E_n(x)$ is the exponential integral of order n. This FIE has a singularity at $\tau'=\tau$, therefor solving this FIE, the method of subtraction of singularity is utilized.\par
	
	The basic state radiative flux is written as
	
	\begin{equation*}
		\boldsymbol{q_s}=\int_0^{4\pi}\left(I_s^c(z,\theta)+I_s^d(z,\theta)\right)\boldsymbol{s}d\Omega=-I_t\exp\left(\int_z^1-\kappa n_s(z')dz'\right)\hat{\boldsymbol{z}}+\int_0^{4\pi}I_s^d(z,\theta)\boldsymbol{s}d\Omega.
	\end{equation*}
	
	since $I_s^d(z,\theta)$ is independent of $\phi$, so x and y components of $\boldsymbol{q_s}$ vanish. Therefore, $\boldsymbol{q}_s=-q_s\hat{\boldsymbol{z}}$, where $q_s=|\boldsymbol{q_s}|$. The mean swimming direction becomes
	
	\begin{equation*}
		<\boldsymbol{p_s}>=-T_s\frac{\boldsymbol{q_s}}{q_s}=T_s\hat{\boldsymbol{z}},
	\end{equation*}
	
	where $T_s=T(G_s).$\par
	The basic cell concentration $n_s(z)$ satisfies,
	
	\begin{equation}\label{34}
		\frac{dn_s}{dz}-V_cT_sn_s=0,
	\end{equation}
	which is supplemented by the cell conservation relation
	\begin{equation}\label{35}
		\int_0^1n_s(z)dz=1.
	\end{equation}
	Eqs.~(\ref{33})-(\ref{35}) constitute a boundary value problem which is solved numerically by using a
	shooting method.
	
	\begin{figure*}[!bt]
		\includegraphics{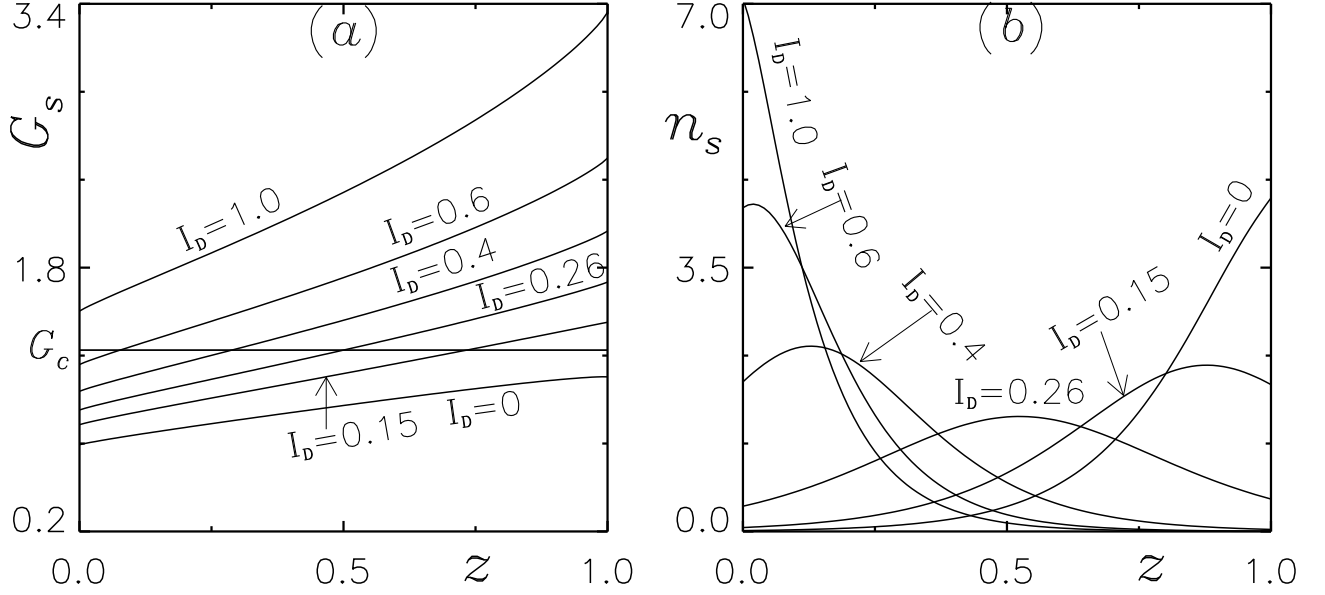}
		\caption{\label{fig3}(a) Variation of total intensity in a uniform suspension for two different values of diffuse irradiation, (b) the corresponding base concentration profile. Here, the  governing parameter values $S_c=20,V_c=10,k=0.5,\omega=0.4$ and $I_t=1$ are kept fixed.}
	\end{figure*}

	We take the incident radiation intensity at the top, $I_t=1.$ Fig.~\ref{fig3} shows the variation of the total intensity $G_s$ across the layer of a uniform suspension $(n=1)$ for $V_c=15,\kappa=0.5$, $\omega=0.4$ and different values of $I_D$.  
	Consider the phototaxis function
	\begin{equation}\label{36}
		T(G)=0.8\sin\left(\frac{3\pi}{2}\chi(G)\right)-0.1\sin\left(\frac{\pi}{2}\chi(G)\right),\quad \chi(G)=\frac{G}{3.8}\exp[0.252(3.8-G)]
	\end{equation}
	with the critical intensity $G_c=1.3$. For $0\leq I_D\leq 1$, $G_s$ is monotonically decreasing across the suspension  (see Fig.~\ref{fig3}(a)). When $I_D=0$, the critical intensity occurs at the top of the domain. Therefore, the maximum concentration occurs at the domain's top. As $I_D$ is increased, the maximum concentration decreases and shifts towards the mid of the domain. At the mid-height of the domain, the maximum concentration is the smallest. As $I_D$ is further increased, the maximum concentration decreases and shifts towards the bottom of the domain (see Fig.~\ref{fig3}(b)). When there is no diffuse irradiation, the uniform total intensity decreases monotonically throughout the suspension for $0<\omega<0.7$. So, the Effect of diffuse irradiation is the same on the basic concentration for $0<\omega<0.7$. \par

	\begin{figure*}[!bt]
		\includegraphics{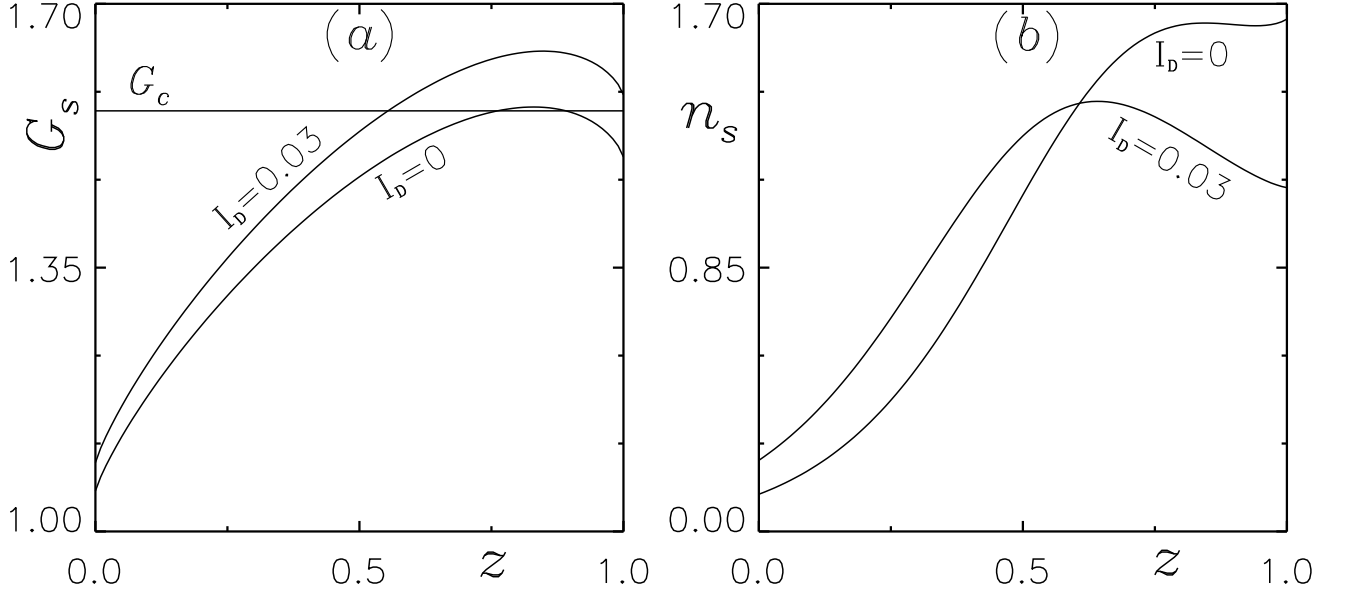}
		\caption{\label{fig4}(a) Variation of total intensity in a uniform suspension for two different values of diffuse irradiation, (b) the corresponding base concentration profile. Here, the  governing parameter values $S_c=20,V_c=10,k=0.5,\omega=1$ and $I_t=1$ are kept fixed.}
	\end{figure*}

	For the case of purely scattering suspension ($\omega=1$), the variation of total intensity and corresponding cell concentration profile is shown in Fig.~\ref{fig4}. In this case, the $G_s$ is not monotonic (i.e., first $G_s$ increases due to scattering and decreases with the depth of the suspension). The taxis function
	\begin{equation}\label{37}
		T(G)=0.8\sin\left(\frac{3\pi}{2}\chi(G)\right)-0.1\sin\left(\frac{\pi}{2}\chi(G)\right),\quad \chi(G)=\frac{1}{3.8}G\exp[0.197(3.8-G)]
	\end{equation}
	
	with the critical intensity $G_c=1.55$ is utilized for this case in which $G_c$ occurs at two locations $z\approx 0.77$ and $z\approx 0.88$ for $I_D=0$. Corresponding the cells accumulate at the two locations $z\approx 0.8$ and the top of the domain. Thus, the cells below $z\approx0.77$ and above $z\approx0.88$ are positively phototactic, and those cells in between are negatively phototactic. As a result, the bimodal steady state occurs for $I_D=0$. But for $I_D=0.03$ the critical intensity, $G_c$, occurs at only one location. As a result, cells accumulate at a single location in the basic state see Fig.~$\ref{fig4}$. Therefore, a bimodal steady state transits to a unimodal steady state as $I_D$ increases.

	\section{Linear stability of the problem}
	For stability analysis, we use linear perturbation theory. Here, the small perturbation of amplitude $\epsilon (0<\epsilon<<1)$ is made to the equilibrium state, according to the following equation
	\begin{widetext}
		
		\begin{equation}\label{38}
			[\boldsymbol{u},n,I,<p>]=[0,n_s,I_s,<p_s>]+\epsilon [\boldsymbol{u}_1,n_1,I_1,<\boldsymbol{p}_1>]+\mathcal{O}(\epsilon^2)=[0,n_s,I_s,<p_s>]+\epsilon [\boldsymbol{u}_1,n_1,I_1^d,<\boldsymbol{p}_1>]+\mathcal{O}(\epsilon^2).  
		\end{equation}
	\end{widetext}
	The perturbed variables are substituted into the Eqs.~(\ref{13})-(\ref16) and linearizing about the equilibrium state by collecting $o(\epsilon)$ terms, gives
	\begin{equation}\label{39}
		\boldsymbol{\nabla}\cdot \boldsymbol{u}_1=0,
	\end{equation}
	where  $\boldsymbol{u}_1=(u_1,v_1,w_1)$.
	\begin{equation}\label{40}
		S_{c}^{-1}\left(\frac{\partial \boldsymbol{u_1}}{\partial t}\right)+\boldsymbol{\nabla} P_{e}+Rn_1\hat{\boldsymbol{z}}=\nabla^{2}\boldsymbol{ u_1},
	\end{equation}
	\begin{equation}\label{41}
		\frac{\partial{n_1}}{\partial{t}}+V_c\boldsymbol{\nabla}\cdot(<\boldsymbol{p_s}>n_1+<\boldsymbol{p_1}>n_s)+w_1\frac{dn_s}{dz}=\boldsymbol{\nabla}^2n_1.
	\end{equation}
	If $G=G_s+\epsilon G_1+\mathcal{O}(\epsilon^2)=(G_s^c+\epsilon G_1^c)+(G_s^d+\epsilon G_1^d)+\mathcal{O}(\epsilon^2)$, then the steady collimated total intensity is perturbed as $I_t\exp\left(-\kappa\int_z^1(n_s(z')+\epsilon n_1+\mathcal{O}(\epsilon^2))dz'\right)$  and after simplification, we get
	\begin{equation}\label{42}
		G_1^c=I_t\exp\left(-\int_z^1 \kappa n_s(z')dz'\right)\left(\int_1^z\kappa n_1 dz'\right)
	\end{equation}
	Similarly, $G_1^d$ is given by 
	\begin{equation}\label{43}
		G_1^d=\int_0^{4\pi}I_1^d(\boldsymbol{ x},\boldsymbol{ s})d\Omega.
	\end{equation}
	Similarly, for the radiative heat flux $q=q_s+\epsilon q_1++\mathcal{O}(\epsilon^2)=(q_s^c+\epsilon q_1^c)+(q_s^d+\epsilon q_1^d)+\mathcal{O}(\epsilon^2)$, we find 
	\begin{equation}\label{44}
		\boldsymbol{q}_1^c=I_t\exp\left(-\int_z^1 \kappa n_s(z')dz'\right)\left(\int_1^z\kappa n_1 dz'\right)\hat{z}
	\end{equation}
	and
	\begin{equation}\label{45}
		q_1^d=\int_0^{4\pi}I_1^d(\boldsymbol{ x},\boldsymbol{ s})\boldsymbol{ s}d\Omega.
	\end{equation}
	Now the expression 
	\begin{equation*}
		-T(G_s+\epsilon G_1)\frac{\boldsymbol{q}_s+\epsilon\boldsymbol{q}_1+\mathcal{O}(\epsilon^2)}{|\boldsymbol{q}_s+\epsilon\boldsymbol{q}_1+\mathcal{O}(\epsilon^2)|}-T_s\hat{\boldsymbol{z}},
	\end{equation*}
	gives the perturbed swimming direction on collecting $O(\epsilon)$ terms
	\begin{equation}\label{46}
		<\boldsymbol{p_1}>=G_1\frac{dT_s}{dG}\hat{\boldsymbol{z}}-T_s\frac{\boldsymbol{q_1}^H}{\boldsymbol{q_s}},
	\end{equation}
	where $\boldsymbol{q}_1^H=[\boldsymbol{q}_1^x,\boldsymbol{q}_1^y]$ is the horizontal component of the perturbed radiative flux $\boldsymbol{q}_1$.
	Now substituting the value of $<\boldsymbol{p_1}>$ from  Eq.~$(\ref{46})$ into Eq.~$(\ref{41})$ and simplifying, we get
	\begin{equation}\label{47}
		\frac{\partial{n_1}}{\partial{t}}+V_c\frac{\partial}{\partial z}\left(T_sn_1+n_sG_1\frac{dT_s}{dG}\right)-V_cn_s\frac{T_s}{q_s}\left(\frac{\partial q_1^x}{\partial x}+\frac{\partial q_1^y}{\partial y}\right)+w_1\frac{dn_s}{dz}=\nabla^2n_1.
	\end{equation}
	On taking the double curl and z-component of the Eq.~$(\ref{47})$ for eliminating the pressure gradient $P_e$ and the horizontal component of $u_1$. Then Eqs.~(\ref{39}), (\ref{40}), and (\ref{47}) is reduced to two equations for $w_1$ and $n_1$ by decomposing these quantity into normal modes such that
	\begin{equation}\label{48}
		w_1=W(z)\exp{(\sigma t+i(lx+my))},\quad n_1=\Theta(z)\exp{(\sigma t+i(lx+my))}.  
	\end{equation}
	The governing equation for perturbed intensity $I_1$ can be written as
	\begin{equation}\label{49}
		\xi\frac{\partial I_1}{\partial x}+\eta\frac{\partial I_1}{\partial y}+\nu\frac{\partial I_1}{\partial z}+\kappa( n_sI_1+n_1I_s)=\frac{\omega\kappa}{4\pi}(n_sG_1+G_sn_1),
	\end{equation}
	subject to the boundary conditions
	\begin{subequations}
		\begin{equation}\label{50a}
			I_1(x, y, z = 1, \xi, \eta, \nu) =0,\qquad (\pi/2\leq\theta\leq\pi,0\leq\phi\leq 2\pi), 
		\end{equation}
		\begin{equation}\label{50b}
			I_1(x, y, z = 0,\xi, \eta, \nu) =0,\qquad (0\leq\theta\leq\pi/2,0\leq\phi\leq 2\pi). 
		\end{equation}
	\end{subequations}
	The form of Eq.~$(\ref{49})$ suggests the following expression for $I_1^d$ 
	\begin{equation*}
		I_1^d=\Psi^d(z,\xi,\eta,\nu)\exp{(\sigma t+i(lx+my))}. 
	\end{equation*}
	From Eqs.~(\ref{42}) and (\ref{43}), we get
	\begin{equation}\label{51}
		G_1^c=\left[I_t\exp\left(-\int_z^1 \kappa n_s(z')dz'\right)\left(\int_1^z\kappa n_1 dz'\right)\right]\exp{(\sigma t+i(lx+my))}=\mathcal{G}^c(z)\exp{(\sigma t+i(lx+my))},
	\end{equation}
	and 
	\begin{equation}\label{52}
		G_1^d=\mathcal{G}^d(z)\exp{(\sigma t+i(lx+my))}= \left(\int_0^{4\pi}\Psi^d(z,\xi,\eta,\nu)d\Omega\right)\exp{(\sigma t+i(lx+my))},
	\end{equation}

	where $\mathcal{G}(z)=\mathcal{G}^c(z)+\mathcal{G}^d(z)$ is the perturbed total intensity.
	
	Now $\Psi^d$ satisfies
	\begin{equation}\label{53}
		\frac{d\Psi^d}{dz}+\frac{(i(l\xi+m\eta)+\kappa n_s)}{\nu}\Psi^d=\frac{\omega\kappa}{4\pi\nu}(n_s\mathcal{G}+G_s\Theta)-\frac{\kappa}{\nu}I_s\Theta,
	\end{equation}
	subject to the boundary conditions
	\begin{subequations}
		\begin{equation}\label{54a}
			\Psi^d( 1, \xi, \eta, \nu) =0,\qquad (\pi/2\leq\theta\leq\pi,0\leq\phi\leq 2\pi), 
		\end{equation}
		\begin{equation}\label{54b}
			\Psi^d( 0,\xi, \eta, \nu) =0,\qquad (0\leq\theta\leq\pi/2,0\leq\phi\leq 2\pi). 
		\end{equation}
	\end{subequations}
	Similarly from Eq.~(\ref{46}), we have
	\begin{equation*}
		q_1^H=[q_1^x,q_1^y]=[P(z),Q(z)]\exp{[\sigma t+i(lx+my)]},
	\end{equation*}
	where
	\begin{equation*}
		P(z)=\int_0^{4\pi}\Psi^d(z,\xi,\eta,\nu)\xi d\Omega,\quad Q(z)=\int_0^{4\pi}\Psi^d(z,\xi,\eta,\nu)\eta d\Omega.
	\end{equation*}
	The linear stability equations become
	\begin{equation}\label{55}
		\left(\sigma S_c^{-1}+k^2-\frac{d^2}{dz^2}\right)\left( \frac{d^2}{dz^2}-k^2\right)W=Rk^2\Theta,
	\end{equation}
	\begin{equation}\label{56}
		\left(\sigma+k^2-\frac{d^2}{dz^2}\right)\Theta+V_c\frac{d}{dz}\left(T_s\Theta+n_s\mathcal{G}\frac{dT_s}{dG}\right)-i\frac{V_cn_sT_s}{q_s}(lP+mQ)=-\frac{dn_s}{dz}W,
	\end{equation}
	subject to the boundary conditions
	\begin{equation}\label{57}
		W=\frac{dW}{dz}=\frac{d\Theta}{dz}-V_cT_s\Theta-n_sV_C\mathcal{G}\frac{dT_s}{dG}=0,\quad at\quad z=0,1.
	\end{equation}
	
	Here, $k=\sqrt{(l^2+m^2)}$ is the overall non-dimensional wavenumber. Eqs.~(\ref{55})-(\ref{53}) form an eigen value problem for $\sigma$ as a function of the dimensionless parameters $V_c,\kappa,\sigma,l,m,R$.
	
	Eq.~(\ref{56}) becomes (writing D = d/dz)
	\begin{equation}\label{58}
		\Gamma_0(z)+\Gamma_1(z)\int_1^z\Theta dz+(\sigma+k^2+\Gamma_2(z))\Theta+\Gamma_3(z)D\Theta-D^2\Theta=-Dn_sW, 
	\end{equation}
	where
	\begin{subequations}
		\begin{equation}\label{59a}
			\Gamma_0(z)=V_cD\left(n_s\mathcal{G}^d\frac{dT_s}{dG}\right)-\iota\frac{V_cn_sT_s}{q_s}(lP+mQ),
		\end{equation}
		\begin{equation}\label{59b}
			\Gamma_1(z)=\kappa V_cD\left(n_sG_s^c\frac{dT_s}{dG}\right)
		\end{equation}
		\begin{equation}\label{59c}
			\Gamma_2(z)=2\kappa V_c n_s G_s^c\frac{dT_s}{dG}+V_c\frac{dT_s}{dG}DG_s^d,
		\end{equation}
		\begin{equation}\label{59d}
			\Gamma_3(z)=V_cT_s.
		\end{equation}
	\end{subequations}
	Introducing the new variable
	\begin{equation}\label{60}
		\Phi(z)=\int_1^z\Theta(z')dz',
	\end{equation}
	the linear stability equations become
	\begin{equation}\label{61}
		\left(\sigma S_c^{-1}+k^2-D^2\right)\left( D^2-k^2\right)W=Rk^2D\Phi,
	\end{equation}
	\begin{equation}\label{62}
		\Gamma_0(z)+\Gamma_1(z)\Phi+(\sigma+k^2+\Gamma_2(z))D\Phi+\Gamma_3(z)D^2\Phi-D^3\Phi=-Dn_sW. 
	\end{equation}
	The boundary conditions become,
	
	\begin{equation}\label{63}
		W=DW=D^2\Phi-\Gamma_2(z)D\Phi-\Gamma_3(z)\frac{dT_s}{dG}\mathcal{G}=0,\quad at\quad z=0,1.
	\end{equation}
	
	The system of Eqs.~(\ref{61}) and (\ref{62}) are of the seventh order, as opposed to the original system, which was the sixth order. Therefore additional boundary conditions are required. So, from Eq.~(\ref{60}), the additional boundary condition is,
	\begin{equation}\label{64}
		\Phi(z)=0,\quad at\quad z=1.
	\end{equation}

	\section{SOLUTION PROCEDURE}
	Numerical solutions to Eqs. (\ref{61}) and (\ref{62}) with the boundary condition (\ref{63}) are obtained using a fourth-order accurate, finite-difference scheme based on NRK iterations~\cite{19cash1980}. By resolving a system of first-order, coupled, nonlinear ordinary differential equations, the NRK technique can solve two-point boundary value problems (ODEs). In order to solve the system of ordinary differential equations using the NRK method, a guess is required. The guess is improved through iterations until the requisite accuracy is attained. This numerical method (the NRK routine) is used to calculate the growth rate, Re$(\sigma)$, or neutral stability curves in the $(k, R)$-plane for a given set of fixed parameters. The initial values of $S_c, V_c, \kappa, k$, and $I_D$ are provided, and the values of W and $\Phi$ are determined by applying sinusoidal variation to W and $\Phi$ or by using previously established numerical results. When a solution is obtained, it may be utilized as an initial guess.
	The neutral curve, $R^{(n)}(k)$ (n = 1, 2, 3,...), has an infinite number of branches and represents a unique solution to the linear stability problem for a specific set of fixed parameters. The branch of the solution where $R$ has its minimum value is the most interesting, and the associated bioconvective solution, i.e., $(k_c, R_c)$, is known as the most unstable solution. The wavelength of the initial disturbance can be calculated using the formula $\lambda_c=2\pi/k_c$.
	To produce solutions, convection cells are stacked one over the other along the depth of the suspension. If there are n convection cells stacked vertically, one on top of the other, in a solution, it is said to be of mode n. In many instances, mostly unstable solution occurs on the $R^{(1)}(k)$ branch of the neutral curve is of mode 1.
	A neutral curve is the locus of points where Re$(\sigma)=0$. If Im$(\sigma)=0 $ on such a curve, then the principle of exchange of stability is valid, and the bioconvective solution is said to be stationary (non-oscillatory). Alternatively, oscillatory solutions exist if Im$(\sigma)\neq 0$. Usually, the oscillatory solution occurs due to competition between the stabilizing and destabilizing processes. If the most unstable solution remains on the oscillatory branch of the neutral curve, then the solution is said to be overstable. When an oscillatory solution occurs, the oscillatory branch of the corresponding neutral curve bifurcates from the stationary branch and exists throughout $k\leq k_b$. 

	\section{NUMERICAL RESULTS}
	We have systematically examined the Effect of magnitude of diffuse irradiation $I_D$. During this examination, the other parameters $S_c,I_t,V_c,\kappa$, and $\omega$ are kept fixed.
	Many parameter values make it difficult to get a full picture throughout the whole parameter domain.
	As a result, we look at how a discrete set of constant parametric values affects the onset of bioconvection. $S_c=20$ and $I_t=1$ are constants utilized throughout the study.
	The parameters for the scattering albedo, extinction coefficient, and cell swimming speed are $\omega\in [0: 1]$, $\kappa= 0.5, 1.0$, and $V_c=10,15,20$.
	
	\subsection{WHEN SELF SHADING IS STRONG}
	
	To study the diffuse irradiation on the onset of bioconvection via weak scattering, the self-shading is considered strong by selecting the lower value of scattering albedo $\omega$. Moreover, self-shading is weak and strong when $\kappa=$ 0.5 and 1, respectively. Here, the critical intensity $G_c=1.3$ is utilized throughout the section.

	\subsubsection{$V_c=15$}
\begin{figure*}[!ht]
		\includegraphics{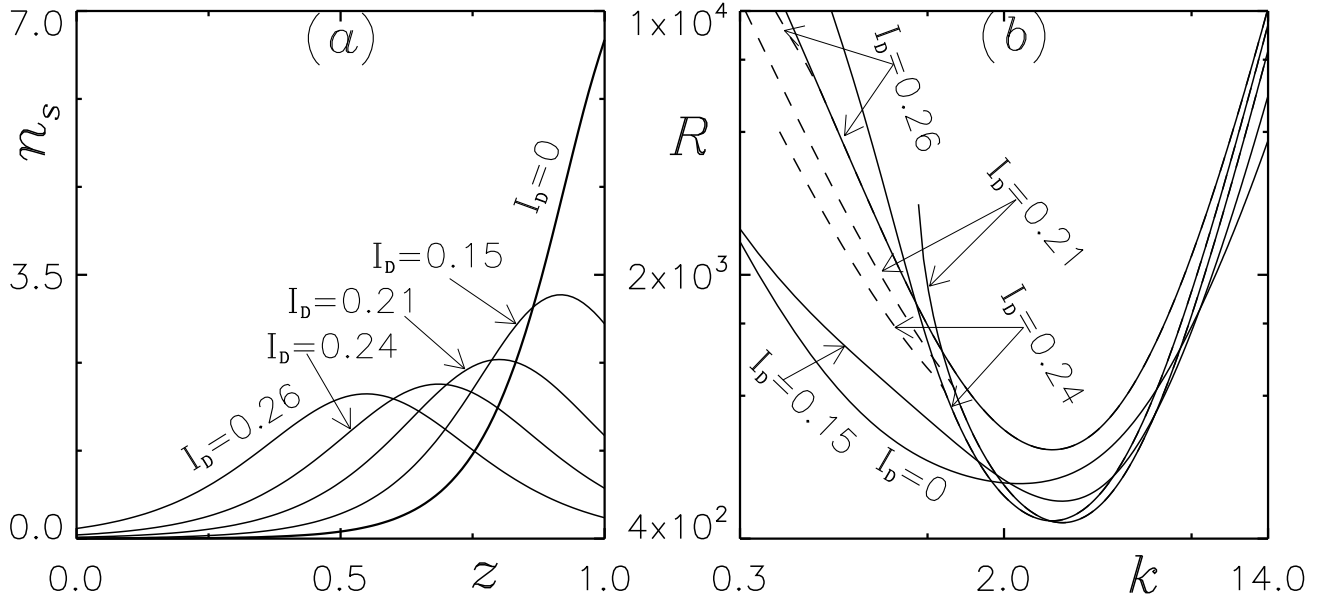}
		\caption{\label{fig5}(a) Basic concentration profile, (b) corresponding neutral curves for different values of $I_D$. Here, the parameter values $S_c=20,V_c=15,k=0.5$, and $\omega=0.4$ are kept fixed.}
	\end{figure*}

	(\romannumeral 1) When extinction coefficient $\kappa=0.5$\\
	The basic cell concentration profile and corresponding neutral curves are shown in fig.~\ref{fig5} for different values of the diffuse light intensity, where the other parameters $v_c=15,k=0.5$, and $\omega=0.4$ are kept fixed. When $I_D=0$, the maximum cell concentration occurs at the top of the domain. As the value of $I_D$ is increased to 0.15, the maximum concentration shifts at $z\approx 0.92$, and the unstable region's width decreases, inhibiting the convective fluid motion. As a result, the lower critical Rayleigh number occurs. The basic state occurs at $z\approx 0.8$ for $I_D=0.21$, and an oscillatory branch bifurcates from the stationary branch of the neutral curve at $k\approx 1.41$ and exists throughout $k\leq 1.41$. But, the most unstable solution occurs on the stationary branch of the neutral curve. As $I_D$ is further increased, at $I_D=0.24$ and 0.26, the maximum concentration occurs at $z\approx 0.69$ and mid-height of the domain, respectively. In these two cases, similar behavior is observed. In all cases, the perturbation to the basic state remains stationary.   
	
	(\romannumeral 2) When extinction coefficient $\kappa=1$\\
	Fig-\ref{fig8} shows the Effect of increment in diffuse irradiation $I_D$ on the basic state and corresponding neutral curves. Here, the governing parameters $V_c=15,\kappa=1,\omega=0.4$ are kept fixed.
	At $_D=0$, the location of the maximum basic concentration is near the top of the domain, and the most unstable bioconvective solution remains in the stationary branch leading the solution to be stationary (non-oscillatory). As $I_D$ increases, the location of the maximum basic concentration
	shifts toward the middle of the chamber. The location of the maximum basic concentration occurs at around $z\approx 0.95$ for $I_D=0.22$, and the corresponding solution is stationary at bioconvective instability. At $I_D=0.38$, the cells accumulate around $z\approx 0.87$ in the basic steady state. In this instance, a single oscillatory branch bifurcates from the basic state at around $k\approx4.59$, but the most unstable solution occurs on the oscillatory branch. So, the perturbation to the basic state is overstable here. When $I_D$ is increased to 0.47, the location of the maximum concentration occurs at around $z\approx 0.73$. In this case, a single oscillatory branch bifurcates from the basic state around $k\approx 3.86$, but the oscillatory branch retains the most unstable solution. Thus, the onset of overstability is at $(k_c, R_c)\approx (2.76,465.54)$. When $I_D$ is further increased to 0.5, the location of the maximum concentration is at the middle of the
	domain, and an oscillatory branch bifurcates from the stationary branch around $k\approx1.06$, exists throughout $k\leq 1.06$. In this instant, the critical Rayleigh number occurs on the stationary branch of the neutral curve. Usually, the critical Rayleigh
	number and wavenumber value increase as ID increases for different fixed governing parameter values. The numerical results for the critical Rayleigh number ($R_c$) and wavenumber ($k_c$) of this section are summarized in Table-1. 
	
	\begin{figure*}[!bt]
		\includegraphics{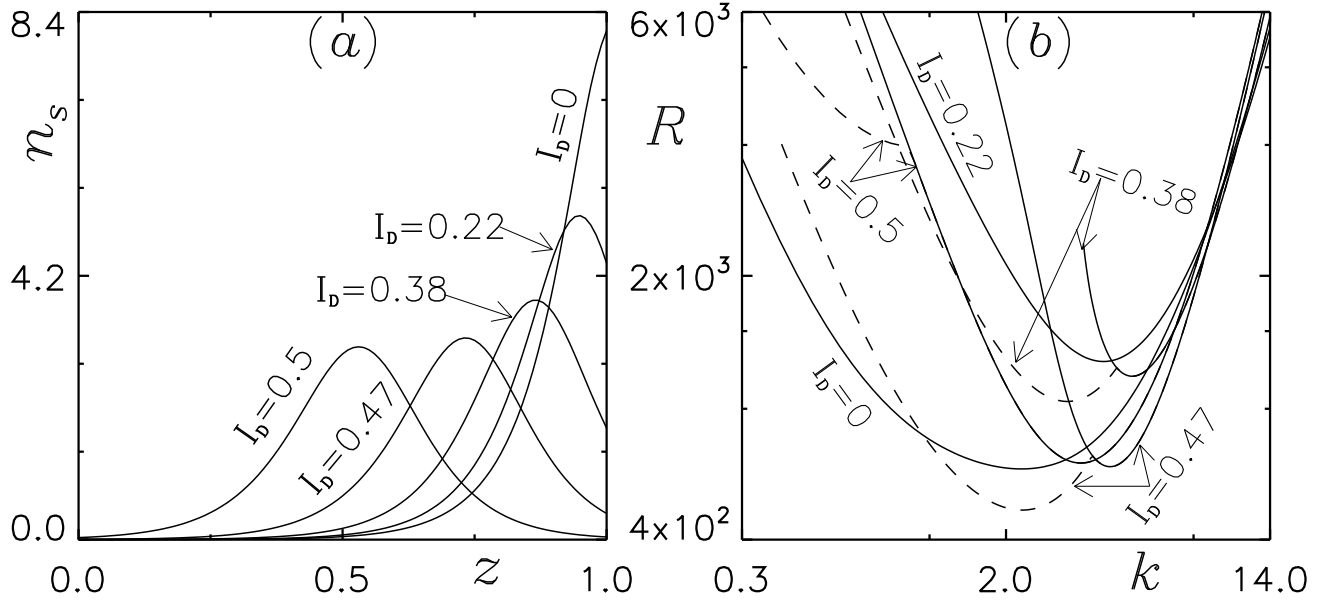}
		\caption{\label{fig6}(a) Basic concentration profile, (b) corresponding neutral curves for different values of $I_D$. Here, the parameter values $S_c=20,V_c=15,k=1$, and $\omega=0.4$ are kept fixed.}
	\end{figure*}

	\begin{table}[h]
		\caption{\label{tab3}The quantitative values of bioconvective solutions with increment in $I_D$ for $V_c=15$ are shown in the table, where other parameters are kept fixed.}
		\begin{ruledtabular}
			\begin{tabular}{ccccccc}
				$V_c$ & $\kappa$ & $\omega$ & $I_D$ & $\lambda_c$ & $R_c$ & $Im(\sigma)$ \\
				\hline
				15 & 0.5 & 0.4 & 0 & 2.74 & 574.99 & 0 \\
				15 & 0.5 & 0.4 & 0.15 & 2 & 512.16 & 0 \\
				15 & 0.5 & 0.4 & 0.21\footnotemark[1] & 1.99 & 444.43 & 0 \\
				15 & 0.5 & 0.4 & 0.24\footnotemark[1] & 2.17 & 449.73 & 0 \\
				15 & 0.5 & 0.4 & 0.26\footnotemark[1] & 2.17 & 719 & 0 \\
				15 & 1 & 0.4 & 0 & 1.95  & 842.82  & 0 \\
				15 & 1 & 0.4 & 0.22 & 1.52  & 997.91 & 0 \\
				15 & 1 & 0.4 & 0.38 & 1.98\footnotemark[2] & 812.89\footnotemark[2] & 17.81 \\
				15 & 1 & 0.4 & 0.47 & 2.76\footnotemark[2] & 465.54\footnotemark[2] & 14.11 \\
				15 & 1 & 0.4 & 0.5\footnotemark[1] & 1.77 & 592.65 & 0 \\
			\end{tabular}
		\end{ruledtabular}
		\footnotetext[1]{A result indicates that the $R^{(1)}(k)$ branch of the neutral curve is oscillatory.}
		\footnotetext[2]{A result indicates that a smaller solution occurs on the oscillatory branch.}
	\end{table}

	\subsubsection{$V_c=10$ and $V_c=20$}
	We have also investigated the Effect of diffuse irradiation on the bio-convective instability for $V_c=10$ and $V_c=20$. Table-2 and Table-3 provide an overview of the numerical data for the critical Rayleigh number and wave number for $V_c=10$ and $V_c=20$.

	\begin{table}[h]
		\caption{\label{tab2}The quantitative values of bio-convective solutions with increment in $I_D$ for $V_c=10$ are shown in the table, where other parameters are kept fixed.}
		\begin{ruledtabular}
			\begin{tabular}{ccccccc}
				$V_c$ & $\kappa$ & $\omega$ & $I_D$ & $\lambda_c$ & $R_c$ & $Im(\sigma)$ \\
				\hline
				10 & 0.5 & 0.4 & 0 & 4.05 & 373.38 & 0 \\
				10 & 0.5 & 0.4 & 0.15 & 2.90 & 390.06 & 0 \\
				10 & 0.5 & 0.4 & 0.2 & 2.88 & 440.17 & 0 \\
				10 & 0.5 & 0.4 & 0.23 & 2.81 & 588.37 & 0 \\
				10 & 0.5 & 0.4 & 0.26 & 2.24 & 1335.27 & 0 \\
				10 & 1 & 0.4 & 0 & 2.76 & 477.94 & 0 \\
				10 & 1 & 0.4 & 0.27 & 1.8 & 576.22 & 0 \\
				10 & 1 & 0.4 & 0.4\footnotemark[1] & 1.73 & 527.75 & 0 \\
				10 & 1 & 0.4 & 0.47\footnotemark[1] & 1.92 & 543.97 & 0 \\
				10 & 1 & 0.4 & 0.5 & 1.92 & 860.48 & 0 \\
				
			\end{tabular}
		\end{ruledtabular}
		\footnotetext[1]{A result indicates that the $R^{(1)}(k)$ branch of the neutral curve is oscillatory.}
		\footnotetext[2]{A result indicates that a smaller solution occurs on the oscillatory branch.}
	\end{table}

	\begin{table}[h]
		\caption{\label{tab4}The quantitative values of bio-convective solutions with increment in $I_D$ for $V_c=20$ are shown in the table, where other parameters are kept fixed.}
		\begin{ruledtabular}
			\begin{tabular}{ccccccc}
				$V_c$ & $\kappa$ & $\omega$ & $I_D$ & $\lambda_c$ & $R_c$ & $Im(\sigma)$ \\
				\hline
				20 & 0.5 & 0.4 & 0 & 2.11 & 910.09 & 0 \\
				20 & 0.5 & 0.4 & 0.19\footnotemark[1] & 1.54 & 673.02 & 0 \\
				20 & 0.5 & 0.4 & 0.24\footnotemark[1] & 1.71 & 478.33 & 0 \\
				20 & 0.5 & 0.4 & 0.25\footnotemark[1] & 1.82 & 444.22 & 0 \\
				20 & 0.5 & 0.4 & 0.26\footnotemark[1] & 1.98 & 478.47 & 0 \\
				
				20 & 1 & 0.4 & 0 & 1.5 & 1483.24 & 0 \\
				20 & 1 & 0.4 & 0.28\footnotemark[1] & 1.04 & 1778.01 & 0 \\
				20 & 1 & 0.4 & 0.41 & 1.84\footnotemark[2] & 924.36\footnotemark[2] & 32.96 \\
				20 & 1 & 0.4 & 0.48 & 2.69\footnotemark[2] & 423.34\footnotemark[2] & 21.52 \\
				20 & 1 & 0.4 & 0.5\footnotemark[1]  & 1.65 & 522.17 & 0 \\
				
			\end{tabular}
		\end{ruledtabular}
		\footnotetext[1]{A result indicates that the $R^{(1)}(k)$ branch of the neutral curve is oscillatory.}
		\footnotetext[2]{A result indicates that a smaller solution occurs on the oscillatory branch.}
	\end{table}
	
	\subsubsection{Effect of swimming speed}
	In this section, we study the Effect of cell swimming speed $V_c$ on the bio-convective instability while the other governing parameters are fixed. Figures 7 to 9  show the basic concentration profile and corresponding neutral curves for three cases $I_D=0.1,0.2,0.26$, where the other governing parameters $\kappa=0.5,\omega=0.4$ are kept fixed. Here, critical intensity $G_c=1.3$ is utilized in all three cases.
	
\begin{figure*}[!bt]
		\includegraphics{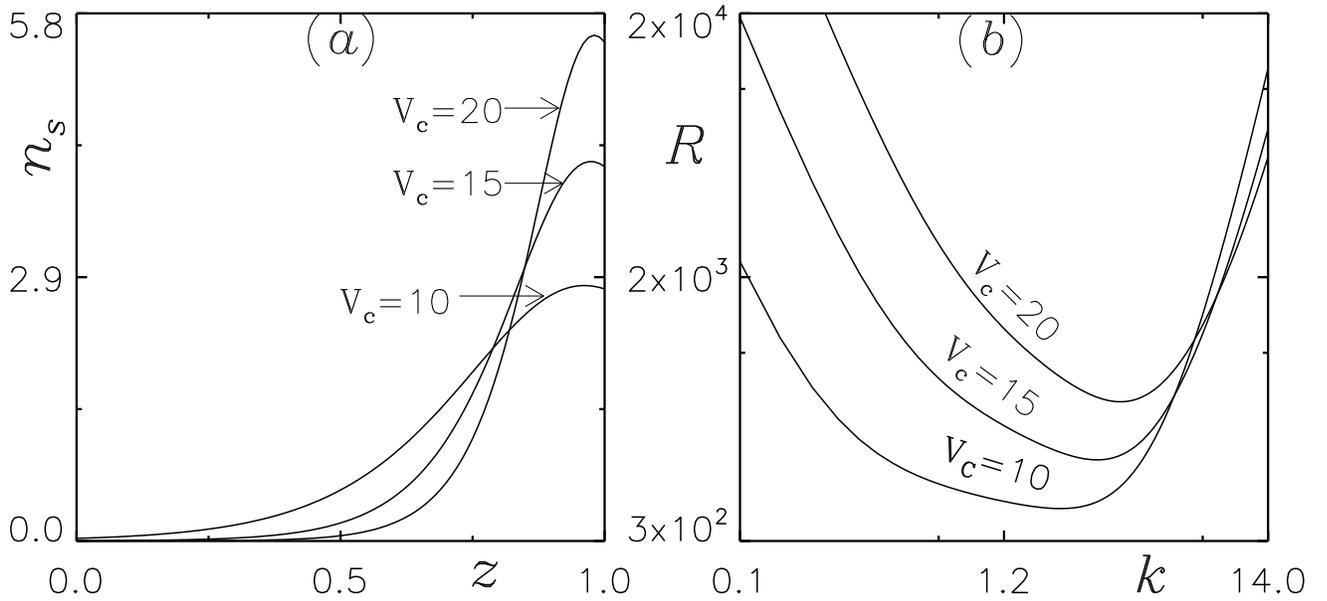}
		\caption{\label{fig7}(a) Basic concentration profiles and (b) corresponding neutral curves for variation in swimming speed $V_c$. Here, other governing parameter values $S_c=20,I_D=0.1,k=0.5$, and $\omega=0.4$ are kept fixed.}
	\end{figure*}

	In the first case, the location of maximum concentration occurs around the top of the suspension, and the maximum concentration becomes steep as $V_c$ increases. The steepness of maximum concentration supports the convective fluid motion.
	At $V_c=10$, the steepness of basic concentration is minimum. As $V_c$ increases to 15,20, the basic concentration becomes steeper, which supports the convection. But at higher swimming speeds, the cells in the plum feel higher resistance due to positive phototaxis. In this case, the latter effect dominates the previous one. As a result, the Rayleigh number decrease as $V_c$ increase, shown in Fig.~\ref{fig7}. In all cases, perturbation to the basic state remains stationary.
	
\begin{figure*}[!bt]
		\includegraphics{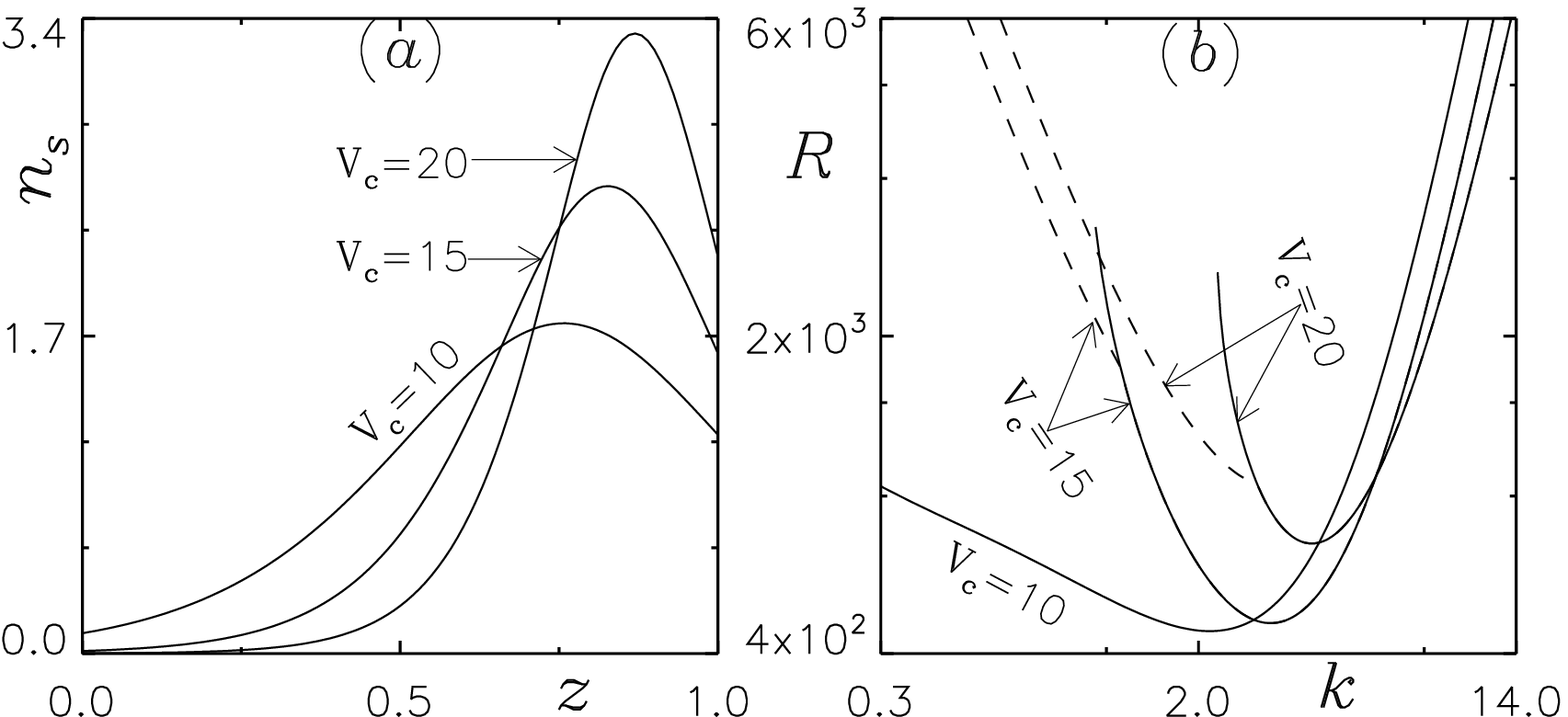}
		\caption{\label{fig8}(a) Basic concentration profiles and, (b) corresponding neutral curves for variation in swimming speed $V_c$. Here, other governing parameter values $S_c=20,I_D=0.2,k=0.5$, and $\omega=0.4$ are kept fixed.}
	\end{figure*}

	In the second case, the effect of cell swimming speed on the basic concentration is shown in Fig.~\ref{fig8}(a), and corresponding neutral stability curves are shown in Fig.~\ref{fig8}(b) at $I_D=0.2$, where other parameters $\kappa=0.5,\omega=0.4$ are kept fixed. In this case, the maximum concentration occurs at $z\approx 0.75$ for $V_c=10$. The maximum concentration increases and moves towards the top of the suspension as $V_c$ increases from 10 to 20. Here, the steepness of concentration increases as $V_c$ increases, but similar to the previous case for $I_D=0.1$, due to positive phototaxis cells in the plum face the higher resistance, which inhibits the convective fluid motion. Therefore, as a result, a higher critical Rayleigh number is observed for higher swimming speed $V_c$. In this case, for $V_c=$ 15 and 20, the oscillatory branch bifurcates from the stationary branch, but the most unstable solution occurs on the stationary branch. Therefore, perturbation to the basic state is stationary here, also.

	In the last case, at $I_D=0.26$, the maximum concentration occurs around the mid-height of the domain for $V_c=10$. As $V_c$ increases, the location of maximum concentration shifts towards the top of the suspension. The steepness of concentration increases with increment in $V_c$, i.e., concentration becomes steeper for $V_c=20$ compared with $V_c=15$ and 10. The steepness of concentration supports the convective fluid motion (convection), but the region of the positive phototaxis, which opposes the convection, increases as an increment in cell swimming speed $V_c$. In this case, the former effect predominates the latter, and a lower critical Rayleigh number is observed. Here also, an oscillatory branch is seen, which bifurcates from the stationary branch of the neutral curve, but the most unstable solution occurs on the stationary branch. Similar to the previous case, perturbation to the basic state is stationary here.

	\begin{figure*}[!bt]
		\includegraphics{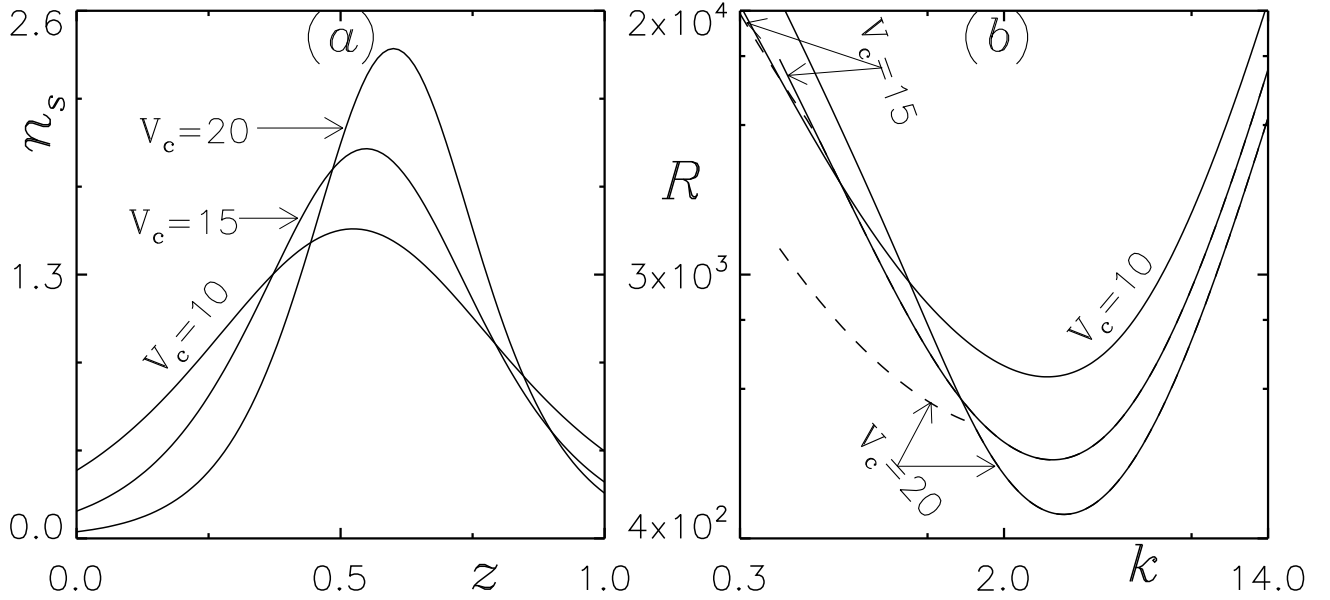}
		\caption{\label{fig9}(a) Basic concentration profiles and (b) corresponding neutral curves for variation in swimming speed $V_c$. Here, other governing parameter values $S_c=20,I_D=0.26,k=0.5$, and $\omega=0.4$ are kept fixed.}
	\end{figure*}
	
	We have also investigated the Effect of cell swimming speed on biocovective solution at absorption coefficient $\kappa=1$. The effects of the cell swimming speed on the bioconvective solutions remain qualitatively similar to the case of $\kappa=0.5$. However, the overstable nature of the bioconvective solution is also observed for some fixed values of $V_c$ at $I_D=0.4$ and 0.48. The numerical results for bioconvective solutions are shown in Table-4.

	\subsection{WHEN SCATTERING IS STRONG}
	
	\begin{figure*}[!bt]
		\includegraphics{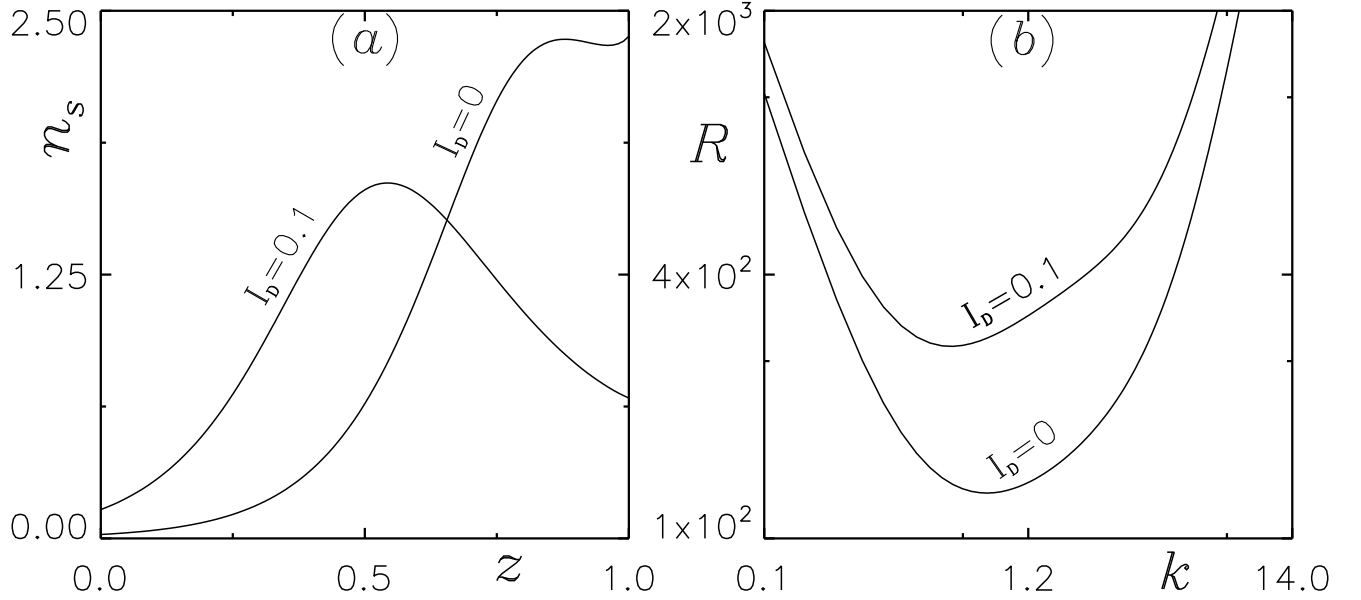}
		\caption{\label{fig10}(a) Basic concentration profiles and (b) corresponding neutral curves for $I_D=0$ and 0.1. Here, the suspension is assumed to be almost purely scattering ($\omega=1$), and other governing parameter values $S_c=20,k=1$, and $V_c=10$ are kept fixed.}
	\end{figure*}

	In this section, we investigate the Effect of diffuse irradiation for purely scattering suspension (i.e., $\omega=1$) on critical wavenumber and Rayleigh number at bioconvective instability by considering the case when self-shading is negligible by taking the purely scattering suspension ($\omega=1$). To emphasize the Effect of diffuse irradiation, we have taken discrete varying parameters as $\kappa=1$, $\omega=1$, and we assume three different cases as $V_c=10,15,20$. In each case, we very magnitude of diffuse irradiation from $I_D=0$ to $I_D=0.1$. Here, we consider the phototaxis function
	\begin{equation}
		T(G)=0.8\sin\left(\frac{3\pi}{2}\chi(G)\right)-0.1\sin\left(\frac{\pi}{2}\chi(G)\right),\quad \chi(G)=\frac{1}{3.8}G\exp[0.135(3.8-G)],
	\end{equation}
	with the critical intensity $G_c=1.9$ in each case.
	
	\begin{figure*}[!bt]
		\includegraphics{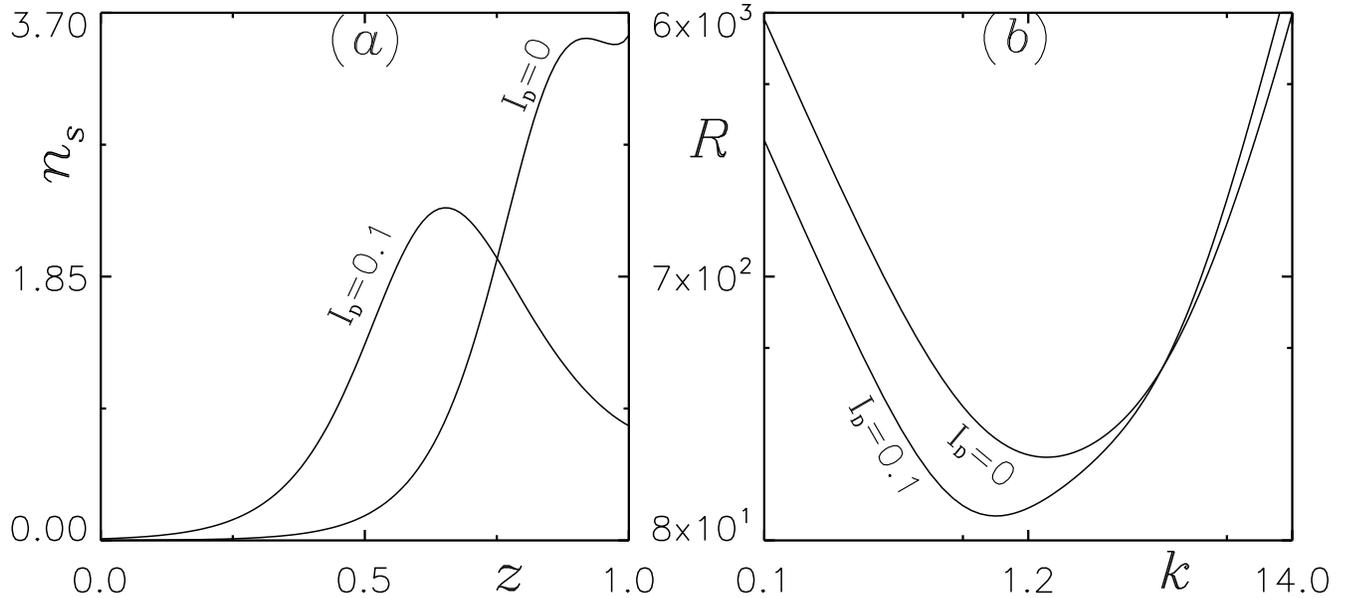}
		\caption{\label{fig11}(a) Basic concentration profiles and, (b) corresponding neutral curves for $I_D=0$ and 0.1. Here, the suspension is assumed to be almost purely scattering ($\omega=1$), and other governing parameter values $S_c=20,k=1$, and $V_c=15$ are kept fixed.}
	\end{figure*}

	Fig.~\ref{fig10}, show the cell basic concentration profile and neutral stability curves for $I_D=0.0.1$ for the case of $V_c=10$, where the other parameters $S_c=20,\omega=1,\kappa=1$ are kept fixed. For $I_D=0$, the bimodal equilibrium state is observed. In this case, the critical intensity ($G_c$) occurs at two locations, $z\approx0.72$ and $z\approx0.88$. Thus below $z\approx0.72$ and above $z\approx0.88$ positive phototaxis occurs, and in between, negative phototaxis occurs, and cells accumulate at two locations at $z\approx0.84$ and top of the suspension (see Fig.~\ref{fig10}). As $I_D$ increases to 0.1, all (some) algae cells between the two locations at $z\approx0.84$ and the top of the suspension swim upward by positive phototaxis due to low light intensity. Thus, the cells accumulate in the basic state at the top of the suspension. Here, the bimodal steady state converts into unimodal. The depth of the unstable region below the maximum concentration in the basic state increases as $I_D$ increases from 0 to 0.1. Since the unstable region supports the convective fluid motion, the critical Rayleigh number increases as $I_D$ increases.
	
	For $V_c=15$, the bimodal steady state is observed in the absence of diffuse irradiation, i.e., $I_D=0$, and this unusual bimodal steady state converts into unimodal as $I_D$ is increased from 0 to 0.1, see Fig.~\ref{fig11}. Here, the depth of the gravitationally stable region increases as the value of the diffuse irradiation is increased which inhibits the convective fluid motion. On the other hand, the maximum concentration becomes steeper for $I_D=0.1$ due to an increment in $V_c$, which supports the convection. The latter effect dominates the former one, and as a result, a lower critical Rayleigh number is observed for $I_D=0.1$ compared to the case of $I_D=0$.
	
\begin{figure*}[!bt]
		\includegraphics{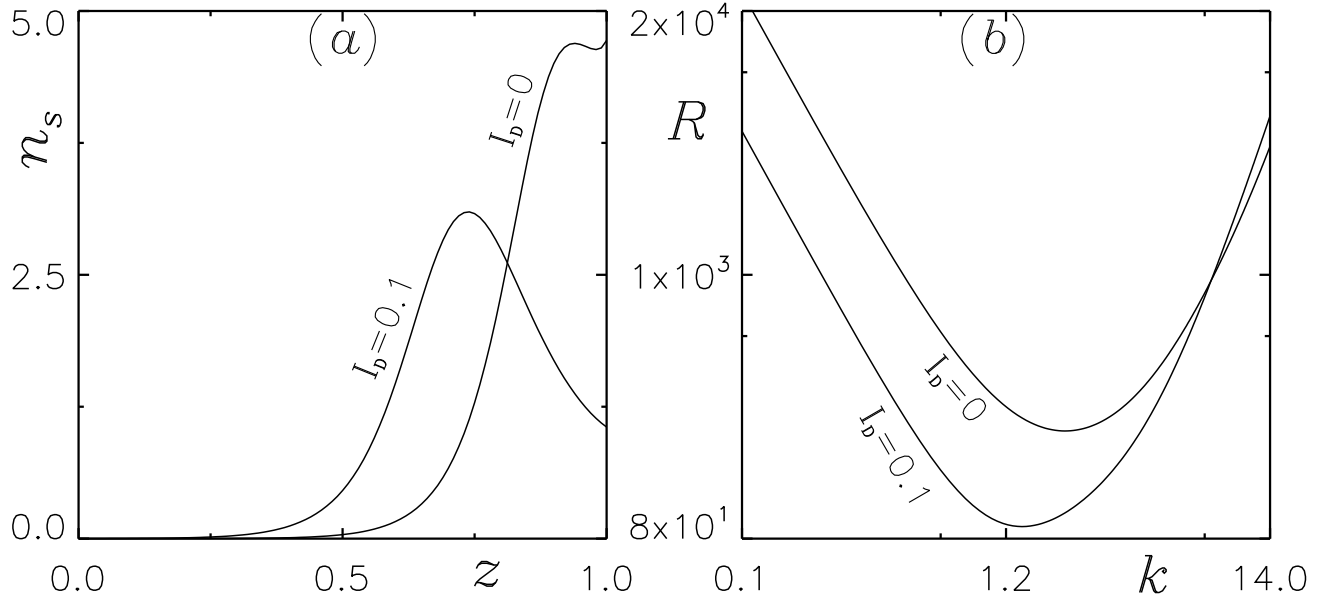}
		\caption{\label{fig12}(a) Basic concentration profiles and (b) corresponding neutral curves for $I_D=0$ and 0.1. Here, the suspension is assumed to be almost purely scattering ($\omega=1$), and other governing parameter values $S_c=20,k=1$, and $V_c=20$ are kept fixed.}
	\end{figure*}

	At $I_D=0$, cells accumulate at two locations, i.e., a bimodal steady state is observed for $V_c=20$. As the value of $I_D$ is increased to 0.1, cells accumulation occurs only in one location in the domain, and a lower critical Rayleigh number is observed for $I_D=0.1$ compared to the case of $I_D=0$, similar to the case of $V_c=15$ (see Fig.~\ref{fig12}).

	\begin{table}
		\caption{\label{tab5}The quantitative values of bioconvective solutions for showing the Effect of variation in cell swimming speed are shown in the table, where other parameters are kept fixed.}
		\begin{ruledtabular}
			\begin{tabular}{ c c c c c c c }
				$I_D$  & $\kappa$ & $\omega$ & $V_c$ & $\lambda_c$ & $R_c$ & $I_m(\sigma)$ \\
				\hline
				
				0.1 & 0.5 & 0.4 & 10 & 3.15 & 381.5  & 0 \\
				0.1 & 0.5 & 0.4 & 15 & 2.21 & 547.53 & 0 \\
				0.1 & 0.5 & 0.4 & 20 & 1.78 & 842.33 & 0 \\
				0.2 & 0.5 & 0.4 & 10 & 2.83 & 440.2 & 0\\			
				0.2 & 0.5 & 0.4 & 15 & 1.96 & 455.47 & 0 \\
				0.2 & 0.5 & 0.4 & 20\footnotemark[1] & 1.54 & 639.57 & 0 \\
				0.26 & 0.5 & 0.4 & 10 & 2.24 & 1335.27 & 0\\
				0.26 & 0.5 & 0.4 & 15 & 2.17 & 719 & 0\\
				0.26 & 0.5 & 0.4 & 20\footnotemark[1] & 1.98 & 478.47 & 0 \\
				0.2 & 1 & 0.4 & 10 & 2.05 & 559.45  & 0\\
				0.2 & 1 & 0.4 & 15 & 1.57 & 978.22 & 0 \\
				0.2 & 1 & 0.4 & 20 & 1.27 & 1658.86 & 0 \\
				0.4 & 1 & 0.4 & 10 & 1.73 & 527.75 & 0 \\
				0.4 & 1 & 0.4 & 15 & 2.05\footnotemark[2] & 724.15\footnotemark[2] & 18.15 \\
				0.4 & 1 & 0.4 & 20 & 1.79\footnotemark[2] & 1002.49\footnotemark[2] & 33.33 \\
				0.48 & 1 & 0.4 & 10 & 1.98 & 592.73 & 0\\
				0.48 & 1 & 0.4 & 15 & 3.02\footnotemark[2] & 459.8\footnotemark[2] & 12.53\\
				0.48 & 1 & 0.4 & 20 & 3.03\footnotemark[2] & 403.39\footnotemark[2] & 11.23 \\
				
			\end{tabular}
		\end{ruledtabular}
		\footnotetext[1]{A result indicates that the $R^{(1)}(k)$ branch of the neutral curve is oscillatory.}
		\footnotetext[2]{A result indicates that a smaller solution occurs on the oscillatory branch.}
	\end{table}

	\begin{table}[h]
		\caption{\label{tab6} The quantitative values of bioconvective solutions for showing the Effect of variation in the value of $I_D$ on purely scattering suspension ($\omega=1$) are shown in the table, where other parameters are kept fixed.}
		\begin{ruledtabular}
			\begin{tabular}{cccccccc}
				$V_c$ & $\kappa$ & $\omega$ & $I_D$ & $\lambda_c$ & $R_c$ \\
				\hline
				10 & 1 & 1 & 0 & 7.77 & 129.36 \\
				10 & 1 & 1 & 0.1 & 10.76 & 297.51 \\
				15 & 1 & 1 & 0 & 4.4 & 158.05 \\
				15 & 1 & 1 & 0.1 & 7.26 & 97.69 \\
				20 & 1 & 1 & 0 & 3.06 & 241.58 \\
				
				20 & 1 & 1 & 0.1 & 4.57 & 90.54 \\

			\end{tabular}
		\end{ruledtabular}
		\footnotetext[1]{A result indicates that the $R^{(1)}(k)$ branch of the neutral curve is oscillatory.}
		\footnotetext[2]{A result indicates that a smaller solution occurs on the oscillatory branch.}
	\end{table}
	
	\section{Conclusion}
	The Effect of a rigid top surface on the onset of phototactic bioconvection in a suspension of isotropic scattering phototactic algae is incorporated in this article. Here, the suspension is illuminated by both collimated and diffuse irradiation from the top. The linear perturbation theory is used to determine the linear stability of the same suspension.
	
	When the suspension's top is free to air, the algae cells form a vertically stacked layer on the top which behaves similarly to a rigid upper surface. This is the effective reason to be assumed a rigid top surface. Due to the presence of the top and bottom rigid surfaces, viscous dissipation reduces, inhibiting the convective fluid motion. As a result, the critical Rayleigh number increases, i.e., a higher Rayleigh number is required for the convective fluid motion which can be seen in numerical results. When the value of diffuse irradiation $I_D$ increases, the Rayleigh number is also increased i.e, Diffuse irradiation and the presence of the rigid vertical surfaces make the suspension more stable.
	
	The Effect of scattering demonstrates no monotonicity in the total intensity variation throughout the suspension depth for certain governing parameter values. This is because the scattering has dual effects when light reaches the microorganisms. First causes the light intensity to drop with the depth of suspension due to the light's deflection away from the incident path at certain locations, and second causing the light intensity to rise due to scattered light arriving at certain locations. As a result, the critical intensity for a greater scattering albedo occurs at two locations in the suspension (for almost purely scattering suspension). Unfortunately, we observe the bimodal steady state, which changes to the unimodal steady state as the value of the diffuse irradiation increases. 
	
	For constant parameter values, the linear stability analysis of the steady state predicts both stationary and oscillatory (overstable) solutions. The oscillatory solution occurs due to conflict between the stabilizing and destabilizing processes. The bioconvective results also show that the stationary solution transits to an oscillatory solution as the diffuse irradiation increases for certain governing parameter values.
	
	However, the suggested phototaxis model should consider comparing the theoretical hypotheses with quantitative experimental results on bioconvection in a purely phototactic algal suspension. Unfortunately, there are currently no such data available.
	So, there is a need to find suitable species of microorganisms that are mainly phototactic, and the most naturally occurring algal species are gravitactic or gyrotactic with phototactic. It's important to note that the suggested model may be used to simulate phototactic bioconvection in an anisotropic algal suspension and several other interesting problems.  
	
	\begin{acknowledgments}
		The author gratefully acknowledges the Ministry of Education (Government of India) for the ﬁnancial support via GATE fellowship (Registration No. MA19S43047204). 
	\end{acknowledgments}
	
	\section*{Data Availability}
	The data that support the plots within this paper and
	other findings of this study are  available
	within the article.
	\nocite{*}
	\bibliography{aipsamp}
	
\end{document}